\documentstyle[12pt,aasms4]{article}

\begin{document}

\title{Matter Mixing in Axisymmetric Supernova Explosion}

\author{Shigehiro Nagataki\altaffilmark{1,2}, Tetsuya M.
Shimizu\altaffilmark{2}, and Katsuhiko Sato\altaffilmark{1,3}}

\noindent
\altaffilmark{1}{Department of Physics, School of Science, University
of Tokyo, 7-3-1 Hongo, Bunkyoku, Tokyo 113, Japan}\\
\altaffilmark{2}{The Institute of Physical and Chemical Research,
Wako, Saitama, 351-01, JAPAN
}\\
\altaffilmark{3}{Research Center for the Early Universe, Faculty of
Science, University of Tokyo, 7-3-1 Hongo, Bunkyoku, Tokyo 113, Japan}


\begin{abstract}

Growth of Rayleigh-Taylor (R-T)
instabilities under the axisymmetric explosion are investigated by
two-dimensional hydrodynamical calculations.
The degree of the axisymmetric explosion and amplitude of the initial
perturbation are varied parametrically to find the most favorable
parameter for reproducing the observed line profile of heavy elements.
It is found that spherical explosion can not produce $\rm ^{56}Ni$
travelling at high velocity ($\sim 3000$km/sec), the presence of which
is affirmed by the observation, 
even if the amplitude of initial perturbation is as large as $30 \%$. 
On the other hand, strong axisymmetric explosion model
produce high velocity $\rm ^{56}Ni$ too much. Weak axisymmetric explosion
are favored for the reproduction of the observed line profile. We
believe this result
shows upper limit of the degree of the axisymmetric explosion.
This fact will be important for the simulation of the collapse-driven 
supernova including rotation, magnetic field, and axisymmetric neutrino
radiation, which have a possibility to cause axisymmetric supernova
explosion.
In addition, the origin of such a large perturbation does not seem to be the
structure of the progenitor but the dynamics of the core collapse
explosion itself since small perturbation can not produce the high
velocity element even if the axisymmetric explosion models are adopted.

\end{abstract}


\section{Introduction} \label{intro}

\indent

It is SN 1987A in Large Magellanic Cloud that provided us 
the apparent evidence of large-scale mixing in the ejecta for the
first  time.
For example, the unexpected early detection of X-rays
(\cite{dotani87}; \cite{sunyaev87}; \cite{wilson88}) and gamma-rays
(\cite{matz88}) suggest the radioactive nuclei,
which are synthesised at the bottom of the ejecta, are mixed up to the 
outer layer.
The form of the X-ray light curve
is also thought to be the indirect evidence of mixing and clumping of the
heavy elements (e.g., \cite{itoh87}; \cite{kumagai88}). 
Moreover, it is reported that a part of heavy elements, such as $\rm
Fe_{II}$, $\rm Ni_{II}$, $\rm Ar_{II}$, and $\rm
Co_{II}$, is mixed up to the fast moving
($\sim$3000-4000 km/s) outer layers from the observation of the width
of its infrared spectral lines (\cite{erickson88}). 
On the other hand, hydrogen which has the
expansion velocity as low as 800 km/s is observed (\cite{hoflich88}). 

At present, the growth of R-T instability 
is thought to be the most promising mechanism for the
explanation of the matter mixing.
Although the idea that the instability grows during
explosion in Type II supernova was not new (e.g., \cite{falk73};
\cite{chevalier76}; \cite{bandiera84}), there was necessity to calculate
numerically the growth of the instability using realistic stellar
models in multi-dimensions to see its effect quantitatively.
With the improvement of the supercomputer, 
many people have done such calculations. Early
two-dimensional simulations of the first few hours of the explosion
showed that the R-T instabilities indeed grows (\cite{arnett89};
\cite{hachisu90}; \cite{muller90}; \cite{fryxell91}; and \cite{muller91}).

However, there are some points which are still open to arguments.
For example, the location and amplitude of the seed of the R-T instability
are still unknown. Up to the present, two candidates have been proposed
for that seed. One is that the convection during the steady stellar
evolution produces such
seed. It is reported that the
density fluctuation, $\delta \rho / \rho$, will be $\sim 5 \%$
(up to $8 \%$) at the
beginning of the core collapse (\cite{bazan94}; \cite{bazan97})
at the inner and outer boundaries of the convective O-- rich shell,
where the radioactive nuclei, such as $\rm ^{56}Ni$, are mainly synthesised. 
The other is that core collapse will amplify the initial fluctuation
in Fe core to the degree of $\delta R_s / R_s \sim 30 \%$, where
$R_s$ is shock radius (\cite{burrows95}).

Another problem is the reproduction of line profiles of heavy
elements.
The line profile of Co and Fe have shown that a small fraction 
of them is expanding at 3000-4000 km/s. On the other hand, numerical
simulations can produce Co and Fe whose velocities are of order 2000
km/s at most even if the acceleration by the energy release of 
the radioactive nuclei is taken into account (\cite{herant91};
\cite{herant92}). Although they insist that pre-mixing of $\rm
^{56}Ni$ will be necessary for the reproduction, the problem of the
high velocity heavy elements seems to be unresolved.

There is another approach to that solution. If
the explosion itself is not spherical symmetry, the situation will change
dramatically. 
There are some reasons we should take account of the asymmetry in
supernova explosion. Among them is a well-known fact that most massive
stars are rapid rotators (\cite{tassoul78}). 
It is well known that stars spin down as they evolve, especially through the 
red supergiant stage, meaning loss of their total angular momentums.
However, pulsars which are found in supernova remnants are
rotating rapidly, to be sure. This fact suggests that a large angular
momentum is still in the center region of the star when it collapses.
Since stars are rotating
in reality, the effect of rotation should be investigated in numerical
simulations of a collapse-driven supernova. Thus far, several
simulations have been done by a few groups in order to study rotating
core collapse (\cite{muller80}; \cite{tohline80}; \cite{muller81};
\cite{bodenheimer83}; \cite{symbalisty84};
\cite{mochmeyer89}; \cite{finn90}; and
\cite{yamada94}).
As a result, some numerical simulations of a collapse-driven supernova
show the possibility of jet-like explosion if the effect of a stellar
rotation and/or stellar magnetic field is taken into
consideration.
There is also a possibility that the axisymmetrically modified neutrino
radiation from a rotating proto-neutron star causes asymmetric explosion
(\cite{shimizu94}). We note these effects mentioned above tend to cause
axisymmetric explosion. 
In addition, axisymmetric explosion has an advantage to the
explanation of some observational facts.
For example, it is reported that axisymmetric explosion has a
possibility to produce
$\rm ^{44}Ti$ so much as to explain the tail of the light curve
of SN 1987A (\cite{nagataki97}).
Furthermore, some observations of SN 1987A suggest
the asymmetry of the explosion. The clearest is the speckle images of
the expanding envelope with high angular resolution
(\cite{papaliolis89}), where an oblate shape with
an axis ratio of $\sim 1.2 - 1.5 $ was shown.  Similar results were also
obtained from the measurement of the linear polarization of the scattered
light from the envelope (\cite{cropper88}). If the envelope is
spherically symmetric, there is no net linear polarization
induced by scattering. Assuming again that the shape of the scattering
surface is an oblate or prolate spheroid, one finds that the observed
linear polarization corresponds to an axis ratio of $\sim 1.2$.
We must note that the observed non-spherical nature of the morphology in
the radio remnant of SN 1987A can be explained by the circumstellar medium
inhomogeneities rather than explosion asymmetry (\cite{gaensler96}). However,
this observation does not necessarily rule out the intrinsic
asymmetric explosion.
Because of these reasons mentioned above, it is important to investigate
the effect of axisymmetric explosion on the mixing of the ejecta.

Based upon these facts, Yamada $\&$ Sato (\cite{yamada91}) did two-
dimensional hydrodynamical calculations
under the axisymmetric and equatorial symmetric explosion. 
They found heavy elements could be highly accelerated in an axisymmetric
explosion and get velocities of order of 4000km/s when the amplitude of
the initial instabilities is as large as $\sim 30\%$.

However, their calculation has some points to be improved as mentioned below.
At first, they calculate matter mixing only with one model, that
is, the initial velocity behind the shock wave is assumed to be
proportional to $r \times \cos^{2} \theta$, where $\theta$ is the zenith
angle. This assumption is groundless and is not persuasive.
Secondly, although they show the presence of the high velocity heavy
elements in the calculation, they do not calculate line profiles of
heavy elements, which should be compared with the observation. 
Thirdly, they assume that the chemical composition of the ejecta and
the mass cut are spherically symmetric. 
Finally, since their numerical algorithm is Donner
Cell method, we feel it necessary to make sure their results by more
refined algorithm.

In this paper, the degree of the axisymmetric explosion is changed
parametrically and the velocity distribution of heavy nuclei
is calculated for each
model. We will make a limit to the degree of the axisymmetric
explosion and the initial fluctuation by comparing the results with
observations.
In each calculation, 
the results of explosive nucleosynthesis under its explosion are used
for the chemical composition and mass cut (\cite{nagataki97}).
Moreover, Roe's  scheme of second-order accuracy in space (\cite{hirsch90};
\cite{shimizu95}; \cite{shimizu96}) is adopted for the calculation.

We show our method of calculation for the matter mixing
in section \ref{calculation}. Results are presented in section
\ref{results}. Summary and discussion are given in section \ref{summary}.

\section{ Model and Calculations } \label{calculation}

\subsection{ Hydrodynamics} \label{hydro}
\indent

We performed two-dimensional hydrodynamical calculations.
The calculated region corresponds to a quarter part of the meridian
plane under the assumption of axisymmetry and equatorial symmetry.
The number of meshes is $ 2000 \times 100$
(2000 in the radial direction, and 100 in the angular direction). 
The size of radial meshes is arranged so as to increase like
geometrical series. Both of the inner most radius and mesh size are set to be
$10^{8} \rm cm$. The outer most radius is set to be $3.3 
\times 10^{12} \rm cm$, that is, the surface of the progenitor.
As for the algorithm, we use the Roe's scheme of a second-order
accuracy in space.
The basic equations are as follows:
\begin{eqnarray*}
\partial_{t} \rho =&& -\frac{1}{r^2}\partial_{r}(\rho u_{r} r^2)
 -\frac{1}{r\sin \theta}\partial_{\theta}(\rho u_{\theta}\sin \theta), \\
\partial_{t} (\rho u_{r}) =&& -\frac{1}{r^2}\partial_{r}(\rho u_{r}^2 r^2)
-\frac{1}{r\sin \theta}\partial_{\theta}(\rho u_r u_{\theta}\sin \theta) \\
&& -\partial_rP + \frac{\rho u_{\theta}^2}{r}, \\
\partial_{t} (\rho u_{\theta}) =&& -\frac{1}{r^2}\partial_r(\rho u_{\theta} u_rr^2) -\frac{1}{r\sin \theta}\partial_{\theta}(\rho u_{\theta}^2\sin \theta)\\
&& -\frac{1}{r}\partial_{\theta}P-\frac{\rho u_{\theta} u_r}{r} ,\\
\partial_tE =&& -\frac{1}{r^2}\partial_r\left[(E+P)u_r r^2\right] \\ 
&& -\frac{1}{\sin \theta}\partial_{\theta}\left[(E+P) u_{\theta}\sin \theta \right]
\\
\end{eqnarray*}
where $\rho, P,$ and $E$ are the mass density,
pressure, total energy
density per unit volume and $u_{r}$ and $u_{\theta}$ are velocities of 
a fluid in $r$ and $\theta$ direction, respectively.
The first equation is the continuity equation, the second and third are 
the Euler
equations and the forth is the equation of the energy conservation.
We use the equation of state:
\begin{eqnarray*}
P = \frac{1}{3} a T^{4} + \frac{\rho k_B T}{ A_{\mu} m_u }
\end{eqnarray*}
where $a$, $ k_B$, $ A_{\mu}$ and $ m_u$
are the radiation constant, Boltzmann
constant, the mean atomic weight, and the atomic mass unit,
respectively.

In this paper, We assume the system is adiabatic after the passage of
the shock wave,
because the entropy produced during the explosive nucleosynthesis is
much smaller than that generated by the shock wave. This means the
effect of nickel bubble is not included in this study.

\subsection{ Post-processing} \label{particle}

\indent

In order to see how the matter is mixed by R-T instabilities quantitatively,
we use a test particle approximation. We will explain this approximation.

At first, test particles are put in the progenitor. 
It is assumed that test particles are at rest and scattered in the
Si-- rich and inner O-- rich layers, where heavy radioactive elements
are mainly synthesised by the explosive nucleosynthesis, with the same interval
in the radial and angular directions in each layer. 
We put $(10(r) \times 100(\theta))$ particles in each layer.
Additionally, we also put $(10(r) \times 100(\theta))$  particles in
outer O--, He--, and H-- rich layers in the same way. The initial positions of
test particles are summarized in Table \ref{position}.

In calculating the degree of the mixing, we assume that each
test particle has its own mass which is determined by the initial 
distribution of the test particles so that their sum becomes 
the mass of the Si--, O--, He--, and H-- rich layers, and also assume 
that each test particle has its own composition which is determined by 
the calculation of explosive nucleosynthesis. We use the results of
\cite{nagataki97} (hereafter NHSY) for the composition as stated below.

As mentioned above, it is assumed that test particles are at rest 
at the beginning (t=0). We also assume that they move with
the local velocity at their positions after the passage of a shock
wave. Thus we can calculate each particle's path by integrating 
$\displaystyle{\partial \vec{x} / \partial t } 
= \vec{v}(t, \vec{x}) $, where
the local velocity $\vec{v}(t, \vec{x})$ is given from the
hydrodynamical calculations.
In this way, we can calculate the degree of the mixing quantitatively 
and can calculate the velocity distribution of each element, such as
$\rm ^{56} Ni$ at each time.

\placetable{position}

\subsection{ Hydrodynamical initial condition} \label{perturbetc}

At first, we will explain the initial shock wave.
Since there is still uncertainty as to the mechanism of Type II supernova,
all calculations of matter mixing have not been performed from
the beginning of the core collapse. 
Instead, 
explosion energy is deposited artificially at the innermost boundary (e.g., 
\cite{hachisu92}). 
In this paper, this method is taken and
the explosion energy of $1.0 \times 10^{51} \rm
erg$ is injected to the region from $1.0 \times 10^{8} \rm cm$ to $1.5
\times 10^{8} \rm cm $ (that is, at the Fe/Si interface).

As for the axisymmetric explosion, 
In this paper, the initial velocity of matter
behind the shock wave is assumed to be
radial and proportional to $r \times [1+\alpha \cos(2 \theta) ] /
 [1+ \alpha ]$, where r, $\theta$, and $\alpha$ are the radius, the zenith
angle, and the free parameter which determine the degree of the
axisymmetric explosion, respectively. 
Since the ratio of the velocity in the polar region to that in the
equatorial region is 1 : $ (1-\alpha)/(1+\alpha)$, 
more extreme jet-like shock waves are obtained as the $\alpha$ gets larger.
In the present study, we take $\alpha=0$ for the spherical explosion 
and $\alpha= 1/3,\; 3/5, \; \rm and \; 7/9$
(these values mean that the ratios of the velocity are 2:1, 4:1, and 8:1,
respectively) for the axisymmetric ones (see Table \ref{model}).
We assumed that the distribution of thermal energy is same as the velocity
distribution and that the total thermal energy is equal to the
total kinetic energy.

Next, in order to see the evolution of the fluctuation, we must introduce
perturbation artificially. 
From the linear stability analysis, the linear growth rate is given as 
(\cite{chandrasekhar81})
\begin{eqnarray*}
G^{2}_{RT} = \frac{\rho _{+} - \rho _{-}}{\rho _{+} + \rho _{-}} kg_{eff}
\end{eqnarray*}
where $\rho _{+}$, $\rho _{-}$, $k$, and $g_{eff}$ are the densities
in the upper and lower layers, the wavenumber of the density
perturbation, and effective gravity, respectively. In our calculation, 
the effective gravity is nearly equal to $- dP/\rho dr $.

As one can see from the growth rate, perturbations of shorter wavelengths
grow faster than that of longer ones. So it is very important
information what the power spectrum of density perturbation of a star
is. 
However, this power spectrum has not been known from observations. 
Only some numerical simulations of the progenitor predict the spectrum 
(\cite{bazan94}; \cite{bazan97}).
Moreover, numerical calculations inherently introduces some amount of
viscosity that suppresses the growth of instabilities of wavelength
shorter than a certain value, which will grow at fastest rate
if such perturbations exist actually. 
We must keep in mind that numerical calculations of R-T instabilities
in a star have uncertainty mentioned above.

Historically speaking, there are two ways of introducing
perturbations. One is periodic method and the other is random
perturbation method. In the periodic perturbation method, a
growing mode is given $a$ $priori$ and characteristic wavelength is
not given in the random perturbation method.
Anyway, their methods assume a form of power spectrum and there is no
guarantee that a real star has such a form.

In this paper, we took a periodic perturbation method since
we consider it better rather than random perturbations since we are
interested in the different growth rate in different direction
(\cite{yamada91}) under the same fluctuation pattern.
As many people have done, we perturb only
velocity field inside the shock wave when the shock front reaches the
He/H interface. When we calculate the axisymmetric explosion, we
introduce perturbation when the shock front in polar region reaches
that interface. We adopt monochromatic perturbations, i.e., $\delta v
\; = \; \varepsilon v(r,\theta) \cos(m \theta)$ (m=20).
In this paper, we performed calculations with three values of
perturbations, that is, $0 \%$, $5 \%$, and $30 \%$ were taken for the 
value of $\varepsilon$. These model parameters are summarized in Table
\ref{models}. We note that it is reported that mixing width, i.e. the
length of the mushroom-like 'fingers', depends
on only slightly on the mesh resolution when $\varepsilon$ is larger
than $\sim 5 \%$ of the expansion speed (\cite{hachisu92}). Because of 
this reason, we think that the influence of the power spectrum of the
initial perturbation is relatively small in this study.

We note that the form of the initial shock wave and the degree of the
initial fluctuation cannot be known directly from both observation and theory.
Rather, we will attempt to make a limit to these initial conditions by
comparing the results with the observations under the assumption of
periodic perturbation.

\placetable{model}
\placetable{models}

\subsection{ Progenitor, Chemical composition, and Mass cut} \label{pro}

\indent

The progenitor of SN 1987A, Sk-69$^{\circ}$202, is thought to
have had the mass $\sim 20 M_{\odot}$ in the main-sequence stage
(\cite{shigeyama88}; \cite{woosley88}) and had $\sim$ (6$\pm$1)$M_{\odot}$
helium core (\cite{woosley88a}).
Thus, we use for the initial density of the progenitor the
presupernova model which is obtained from the evolution of a helium
core of 6 $M_{\odot}$ (\cite{nomoto88}), and their
hydrogen envelope. Total mass of the progenitor corresponds to 16.3
$M_{\odot}$.

\begin{figure}
\epsscale{1.0}
\plottwo{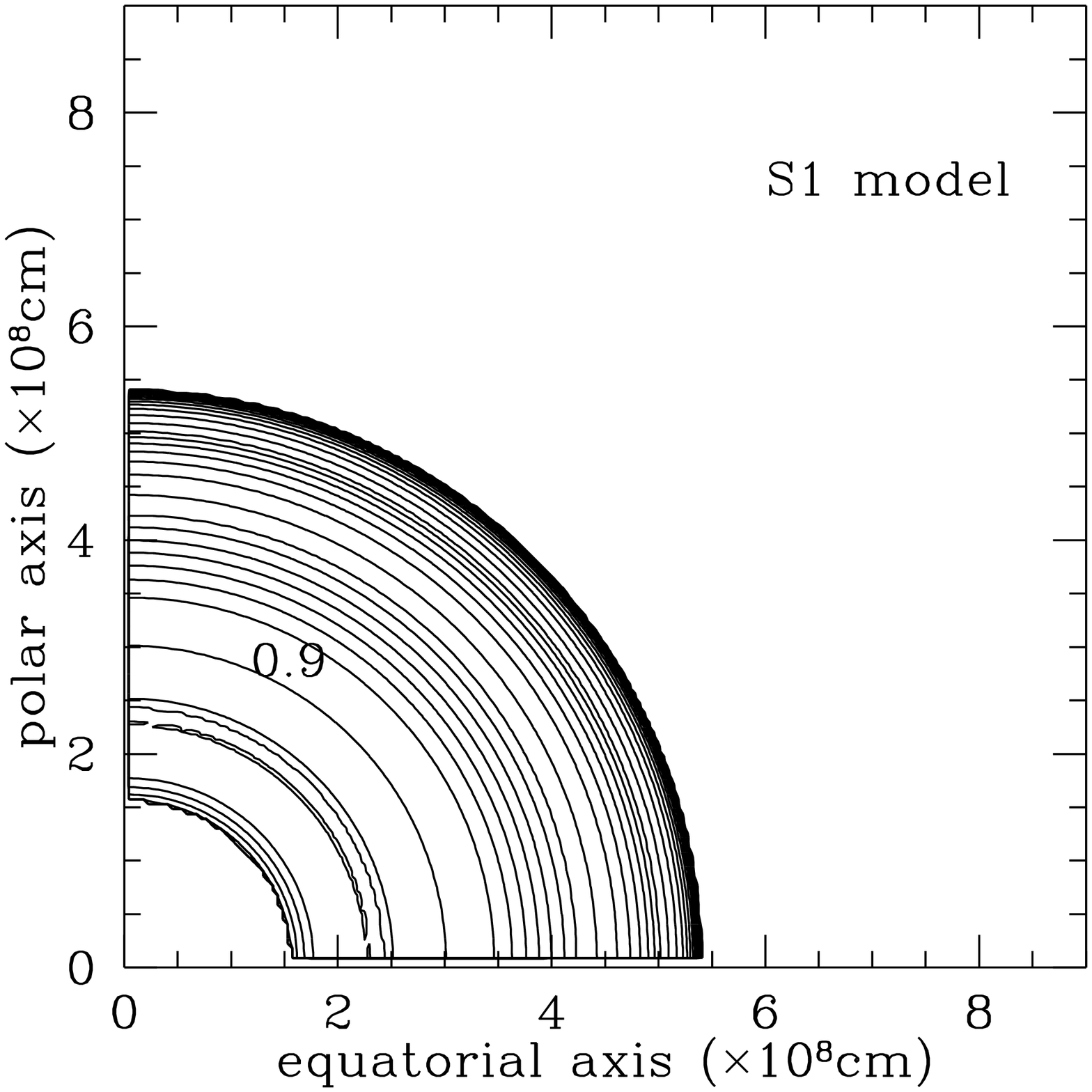}{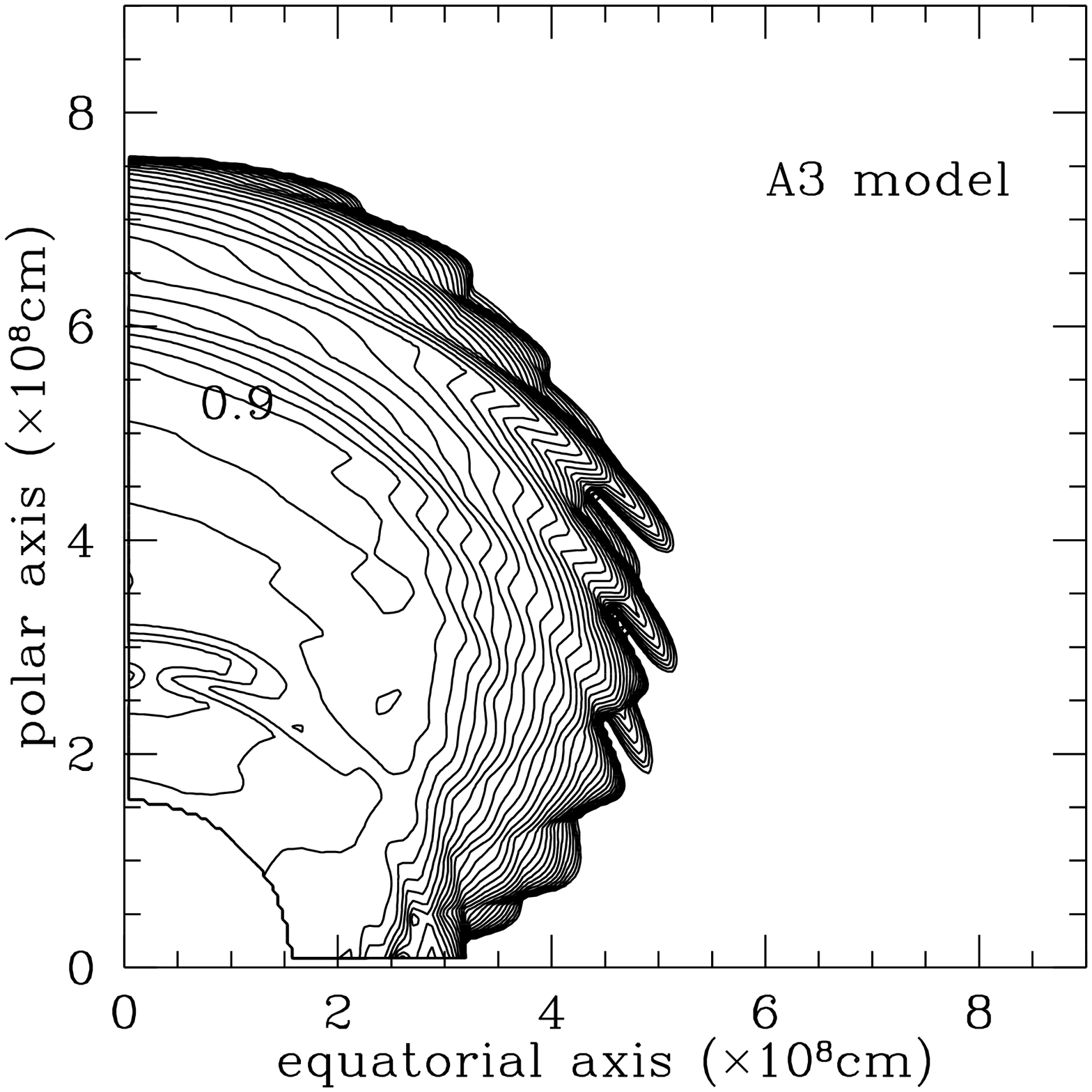}
\figcaption{Left: Contaur of the mass fraction of $^{56}\rm Ni$ in the
S1 model. The maximum value of the mass fraction of$^{56}\rm Ni$ is
9.3$\times 10^{-1}$. The region where mass fraction of $^{56}\rm Ni$
becomes 0.9 is noted in the figure. Right : Same as left but for A3 model. The
maximum value is 9.1$\times 10^{-1}$.   
Contours are drawn for the initial position of test particles.
 \label{fig1}}
\end{figure}

The chemical composition of the progenitor is changed by the nuclear
burning when the shock wave passes (explosive nucleosynthesis). 
We used the results of NHSY for the explosive nucleosynthesis. 
For example, we show in Figure~\ref{fig1} the contour of mass
fraction of $\rm ^{56}Ni$ for S1 and A3 models.  
We note that this effect of the initial asymmetric chemical
composition has never been considered in the previous studies.

\begin{figure}
\epsscale{1.0}
\plottwo{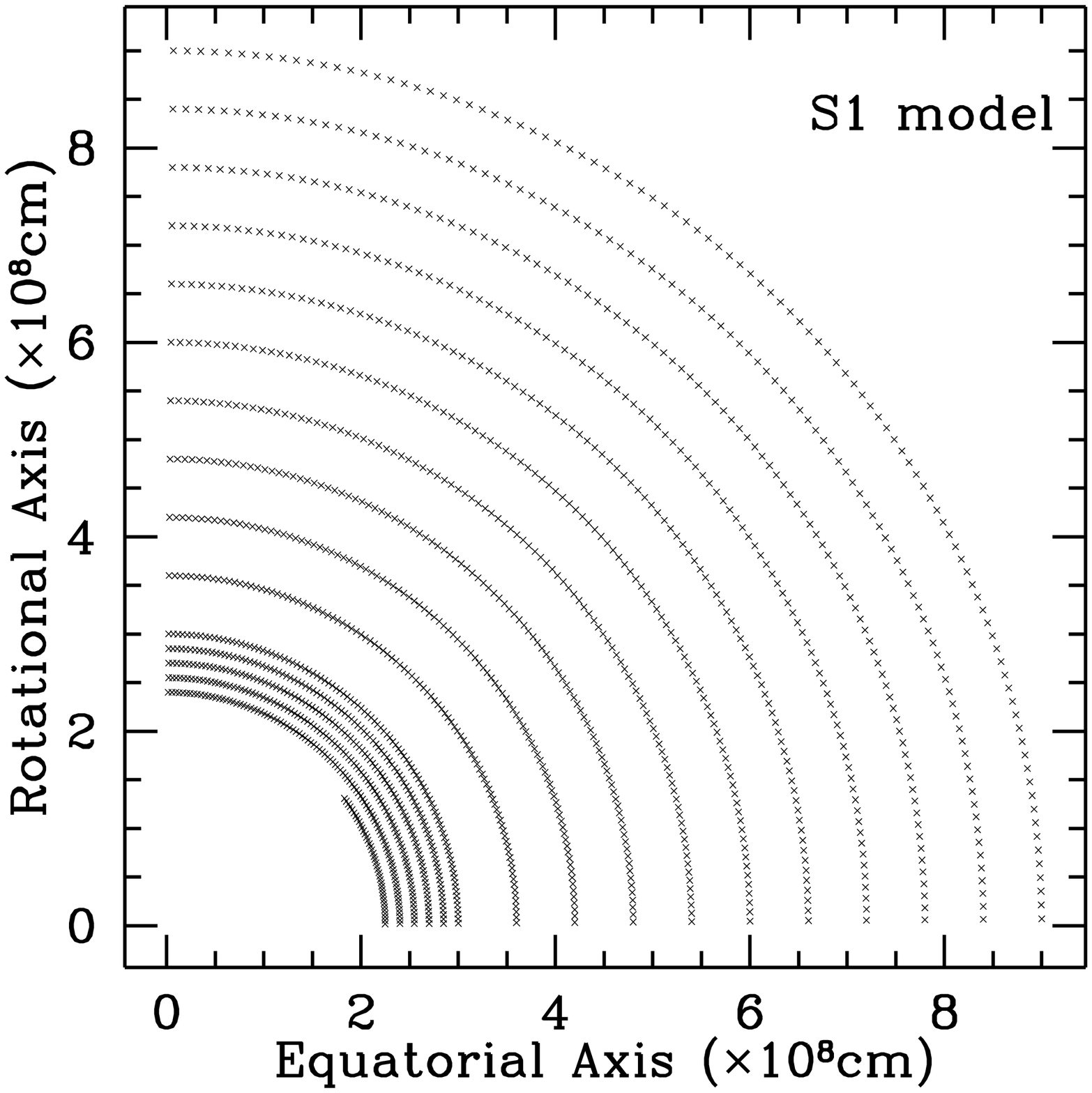}{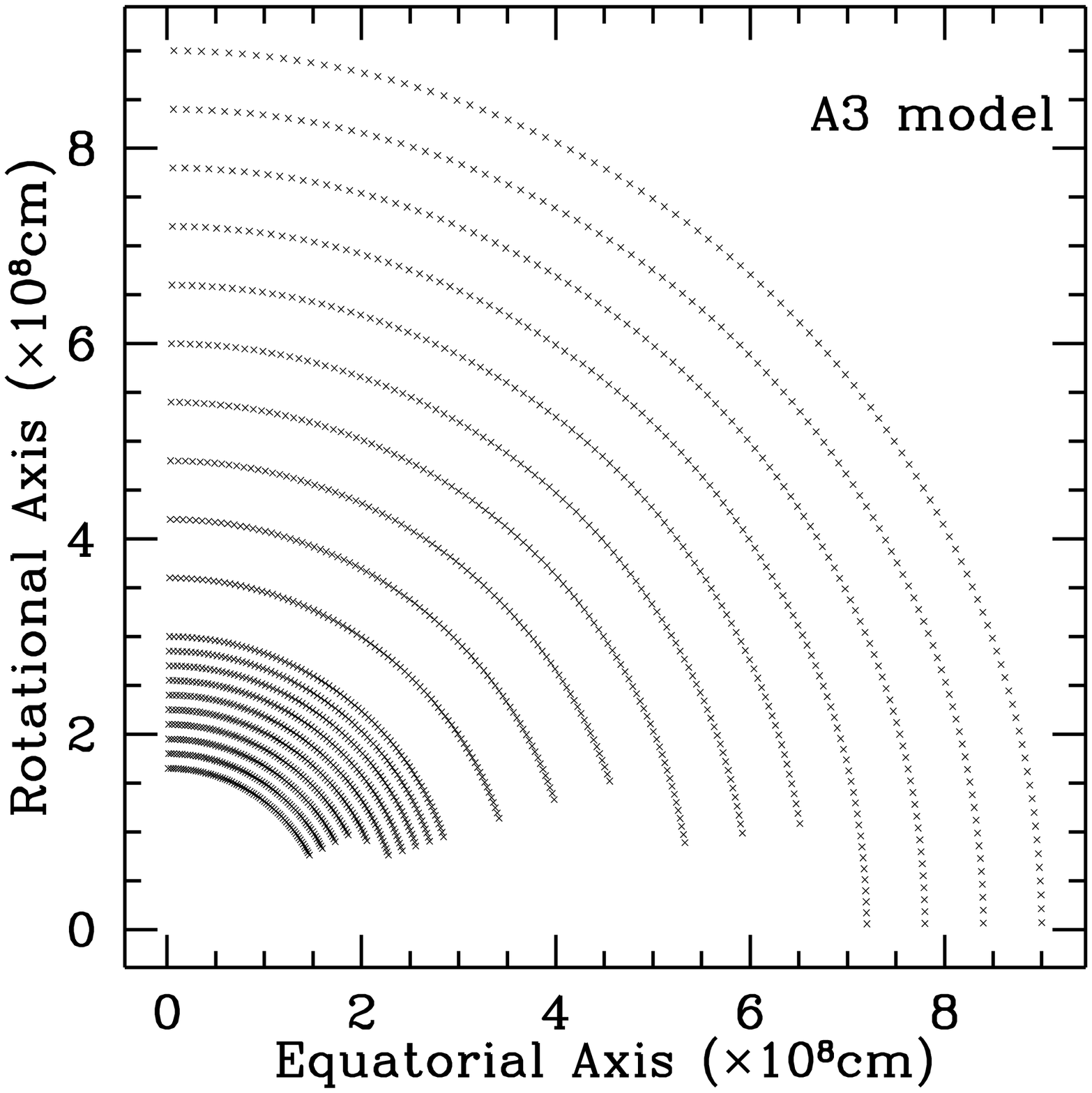}
\figcaption{Form of the mass cut for S1 case and A3 case. 
Dots are plotted for the initial positions of the test particles which 
will be ejected.
\label{fig2}}
\end{figure}

As for the mass cut, we use the results of NHSY. 
We show in Figure~\ref{fig2} the mass cuts for S1 and A3 models.
These mass cuts are chosen so as to contain 0.07$M_{\odot}$ $\rm
^{56}Ni$ in the ejecta. 
However, we must note that defining the form of the mass cut is very
difficult problem since it is sensitive not only to the explosion
mechanism, but also to the presupernova structure, stellar mass, and
metallicity (\cite{woosley95}).
Moreover, the form of the mass cut has a large influence on the mixing of
$\rm ^{56}Ni$ since it is mainly produced near the mass cut. 
Because of these reasons, we must investigate the
sensitivity of the results on the mass cut as seen in section \ref{ni56}.

\placefigure{fig1}
\placefigure{fig2}

\section{ Results} \label{results}

\subsection {Global density structure and Position of $\rm ^{56}Ni$}

\indent

\begin{figure}
\epsscale{0.8}
\plotone{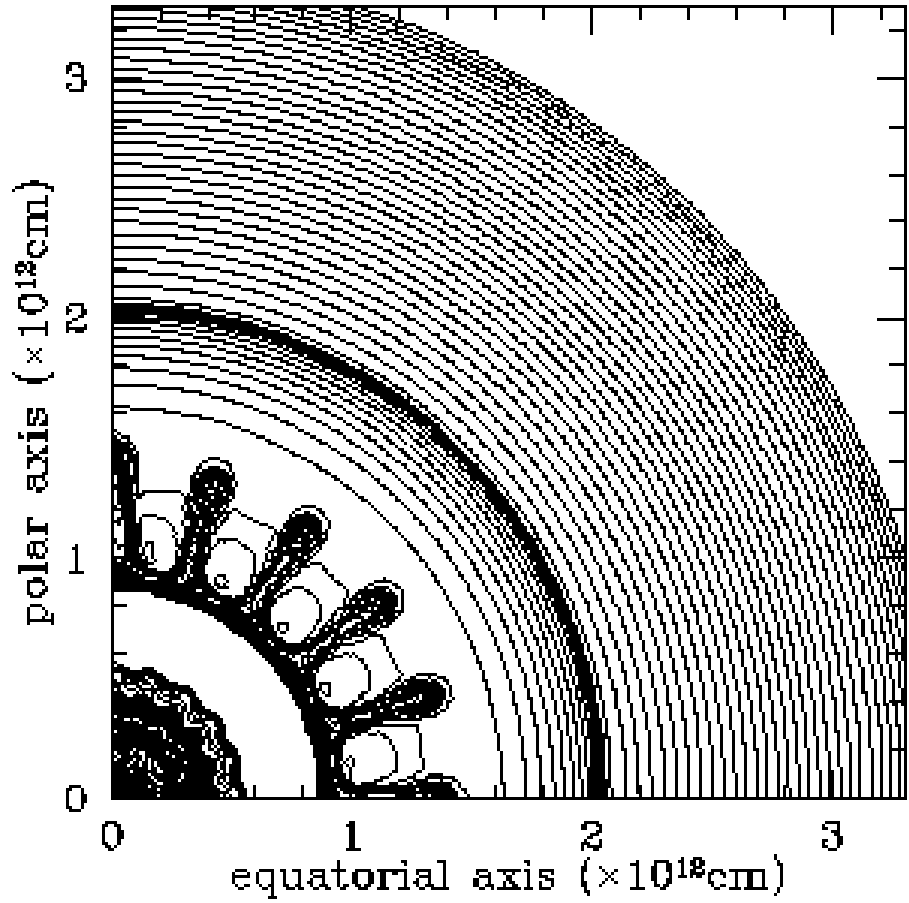}
\figcaption{Density contour for S1b model at t = 5000 sec. The shock
wave can be seen at the radius $\sim 2 \times 10^{12}$ cm. The growth
of R-T instabilities can be seen inside the shock wave. \label{fig3}}
\end{figure}

In this paper, 12 models are performed in all.
We show the density contours of S1b at time = 5000 sec
in Figure~\ref{fig3}.
Mushroom-like 'fingers', which are characteristic of the nonlinear
growth of Rayleigh-Taylor instability, can be seen in this model.
This is consistent with the work done by other groups
(e.g., \cite{fryxell91}; \cite{hachisu92}). From this point of view,
we think that the resolution of our calculations are high enough to
see the global behavior of the matter mixing.

\begin{figure}
\epsscale{0.8}
\plottwo{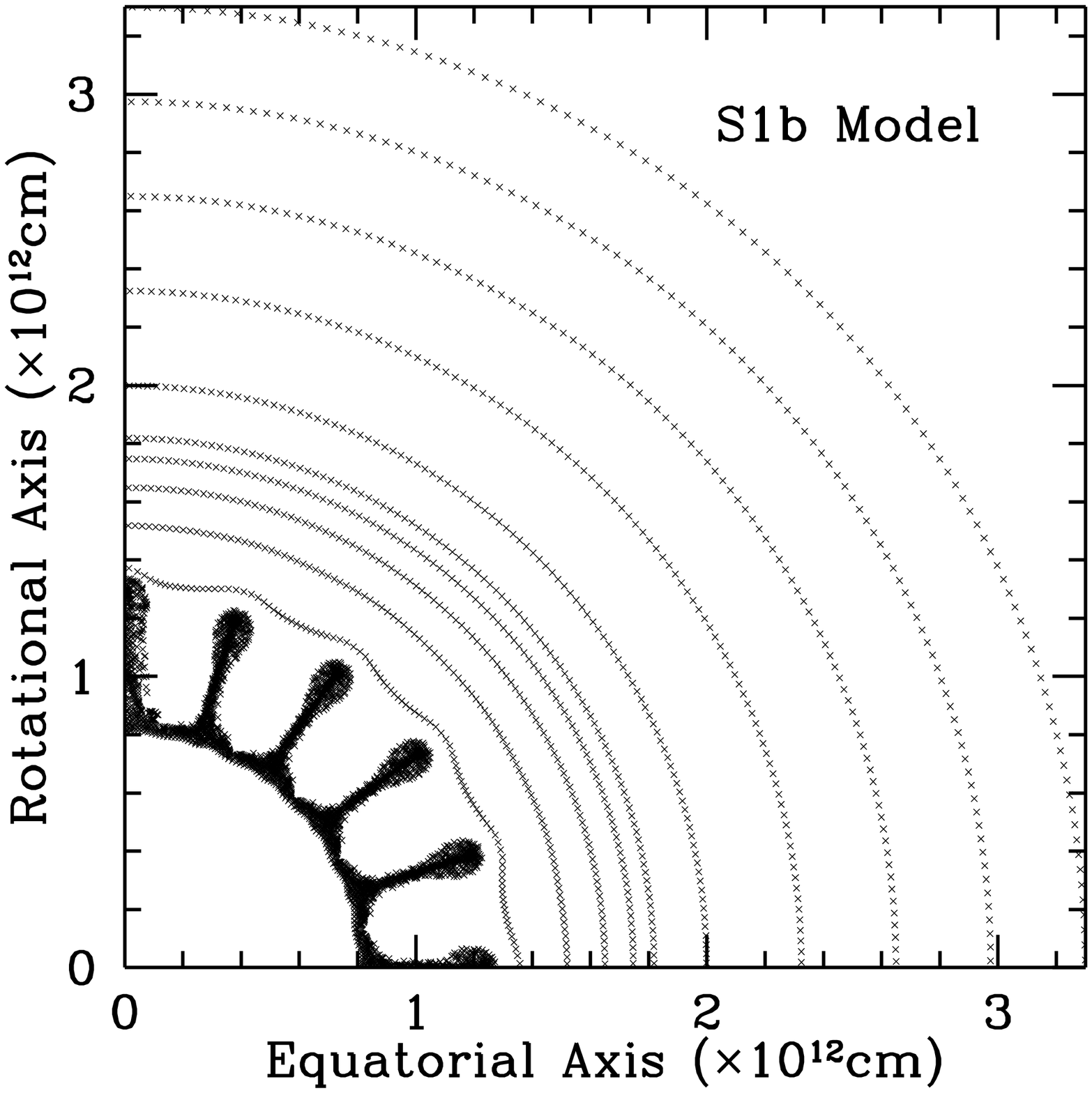}{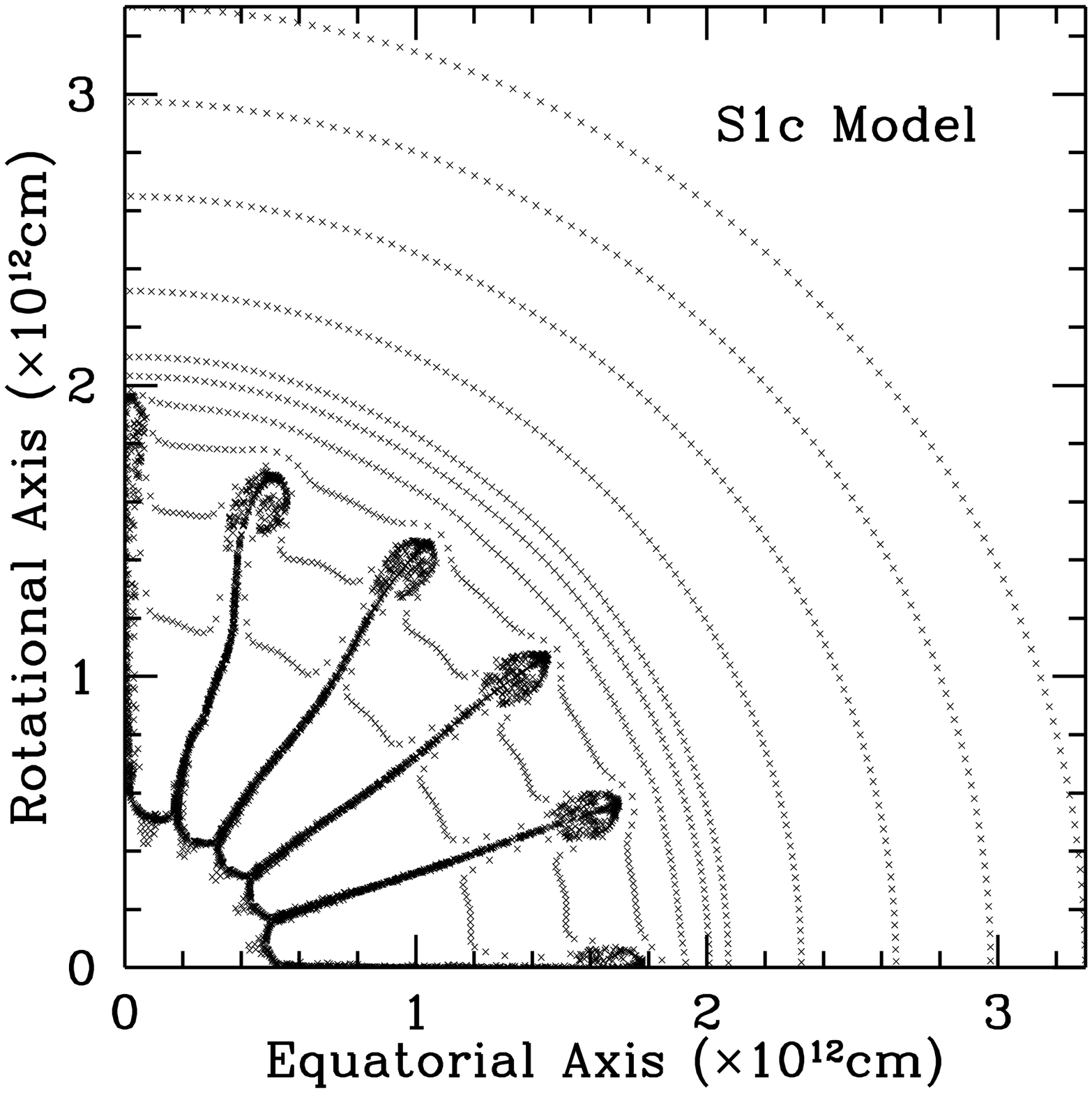}
\figcaption{Positions of test particles at time = 5000 sec after
explosion. Left: S1b model. Right: S1c model.
\label{fig4}}
\end{figure}

\begin{figure}
\epsscale{0.8}
\plottwo{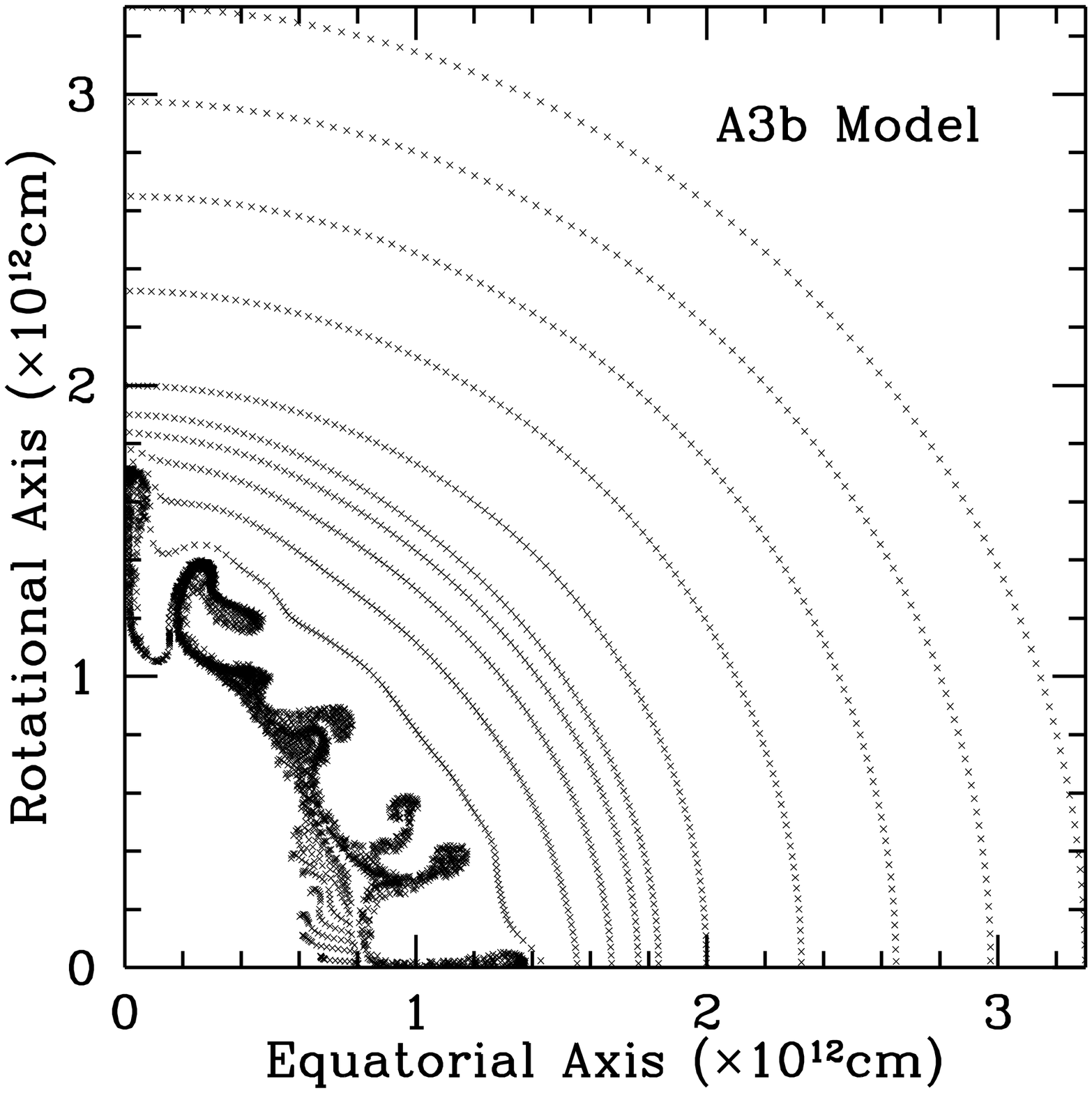}{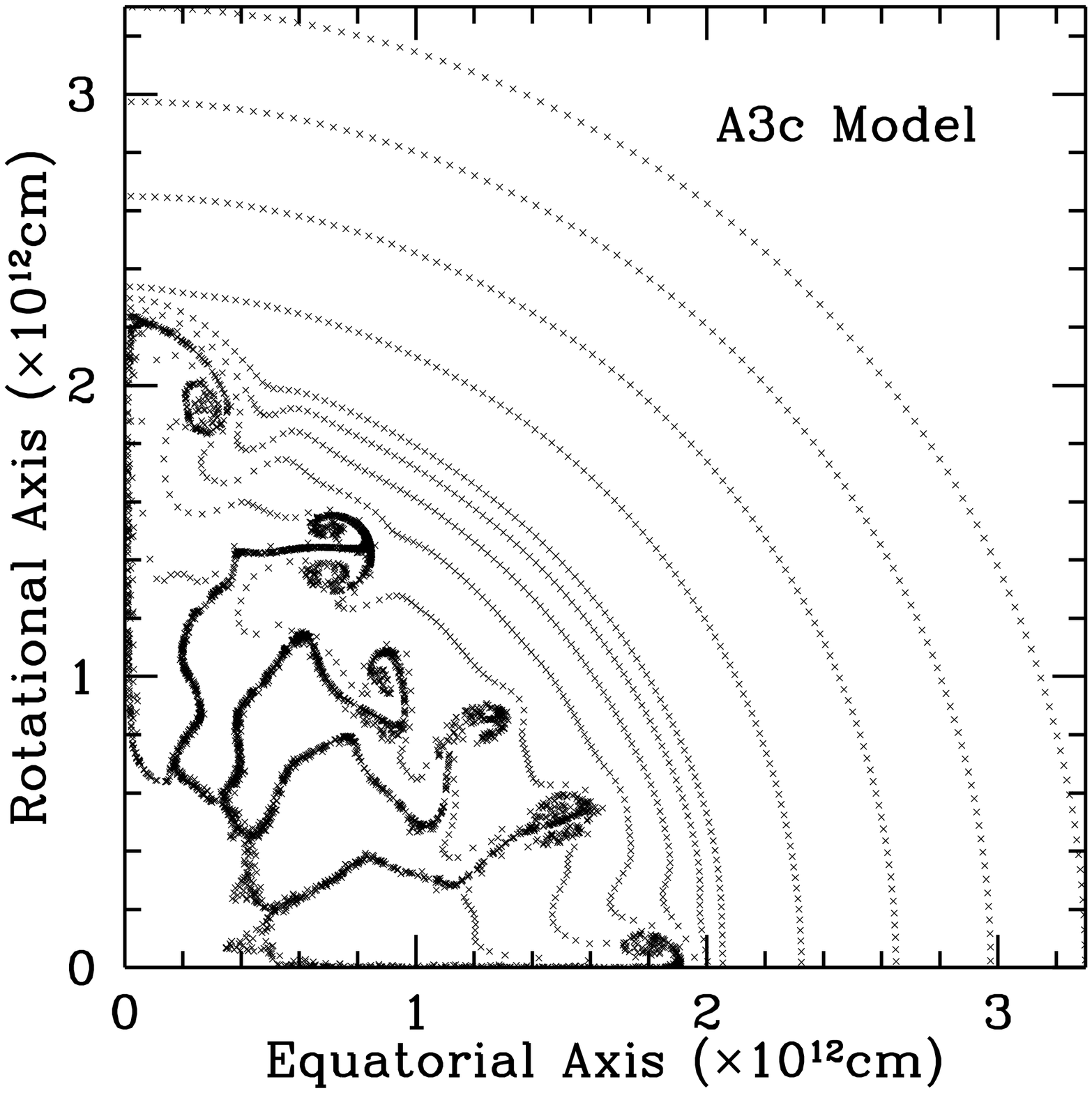}
\figcaption{Same as Fig.4 but for A3b and A3c models.
\label{fig5}}
\end{figure}

Next, positions of test particles are shown at time = 5000 sec for S1b
and S1c in
Figure~\ref{fig4}. For comparison, we show those for A3b and A3c in
Figure~\ref{fig5}. As expected, the matter is mixed more in the
larger initial perturbation model (S1c and A3c models).
We can also see that the matter is mixed more in the polar region than
in the equatorial region for the axisymmetric explosion model (A3b and
A3c models).

Before we go to the next subsection, we must comment on
the effect of two-dimensional calculations.
All calculations presented here are performed in two dimensions
so as to save our computer's CPU-time and memory and to allow the
extensive exploration of the parameter space carried out in this study.
Readers should note here that there is a tendency that
three-dimensional modes grow to higher mixing velocities than
two-dimensional modes (\cite{remington91}; \cite{yabe91}; \cite{marinak95}).
This seems to be the reason why the finger at the $\theta = 0^{\circ}$ 
axis is further along than at
$\theta = 90^{\circ}$ axis even if the spherical explosion model of
S1c. However, three-dimensional simulations of the explosion of SN 1987A
(\cite{muller90}) have shown little change in relation to the two
dimensional case.
Moreover, by comparing figure~\ref{fig4} with figure~\ref{fig5},
we think that the
effect of axisymmetric explosion is quite large compared with the
two-dimensional effect. It is true that the three-dimensional effect
should be investigated with high resolution in the future, but we 
think our quantitative estimates discussed below are not unreasonable.

\placefigure{fig3}
\placefigure{fig4}
\placefigure{fig5}		
					 
\subsection { Velocity distribution of $\rm ^{56}Ni$ } \label{ni56}

\indent 

We will pay attention to $\rm ^{56}Ni$, which is responsible for the
behavior of the bolometric light curve, the early detection of X-yay and
Gamma-ray, and their light curves.

\begin{figure}
\epsscale{0.8}
\plottwo{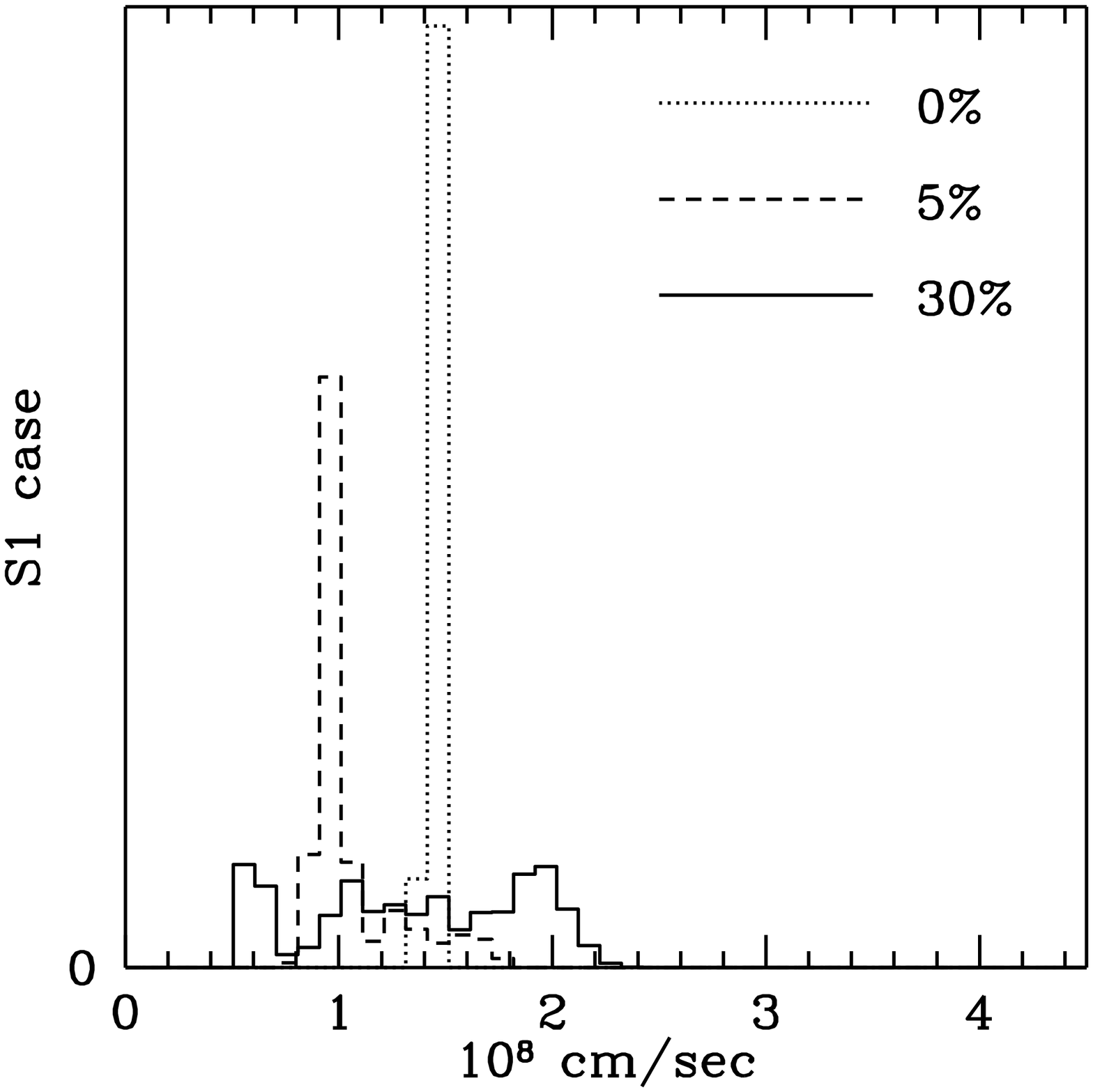}{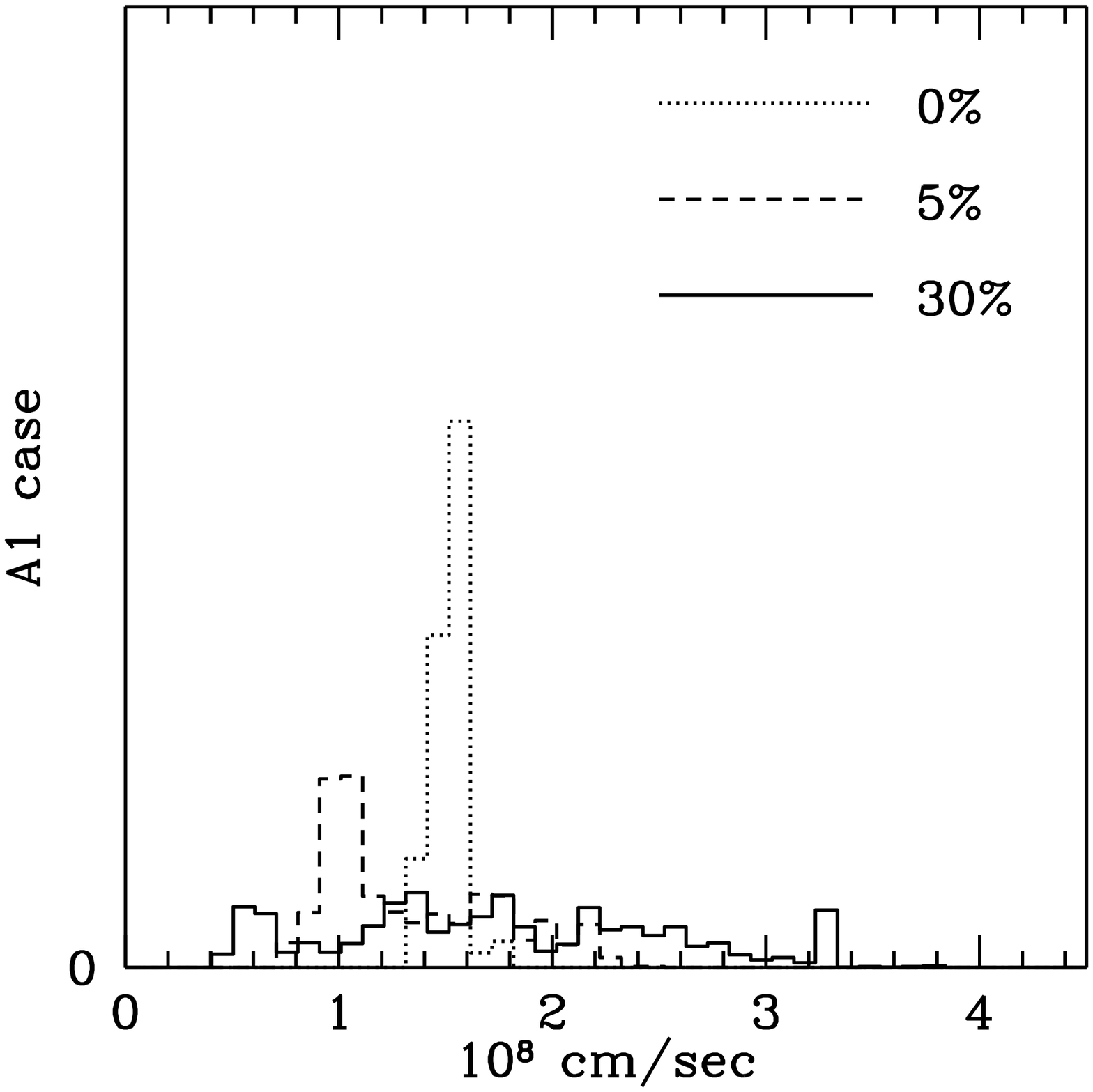}
\figcaption{Velocity distribution of $\rm ^{56}Ni$ at time = 5000 sec
after explosion. Left: S1 model. Right: A1 model. 
\label{fig6}}
\end{figure}

\begin{figure}
\epsscale{0.8}
\plottwo{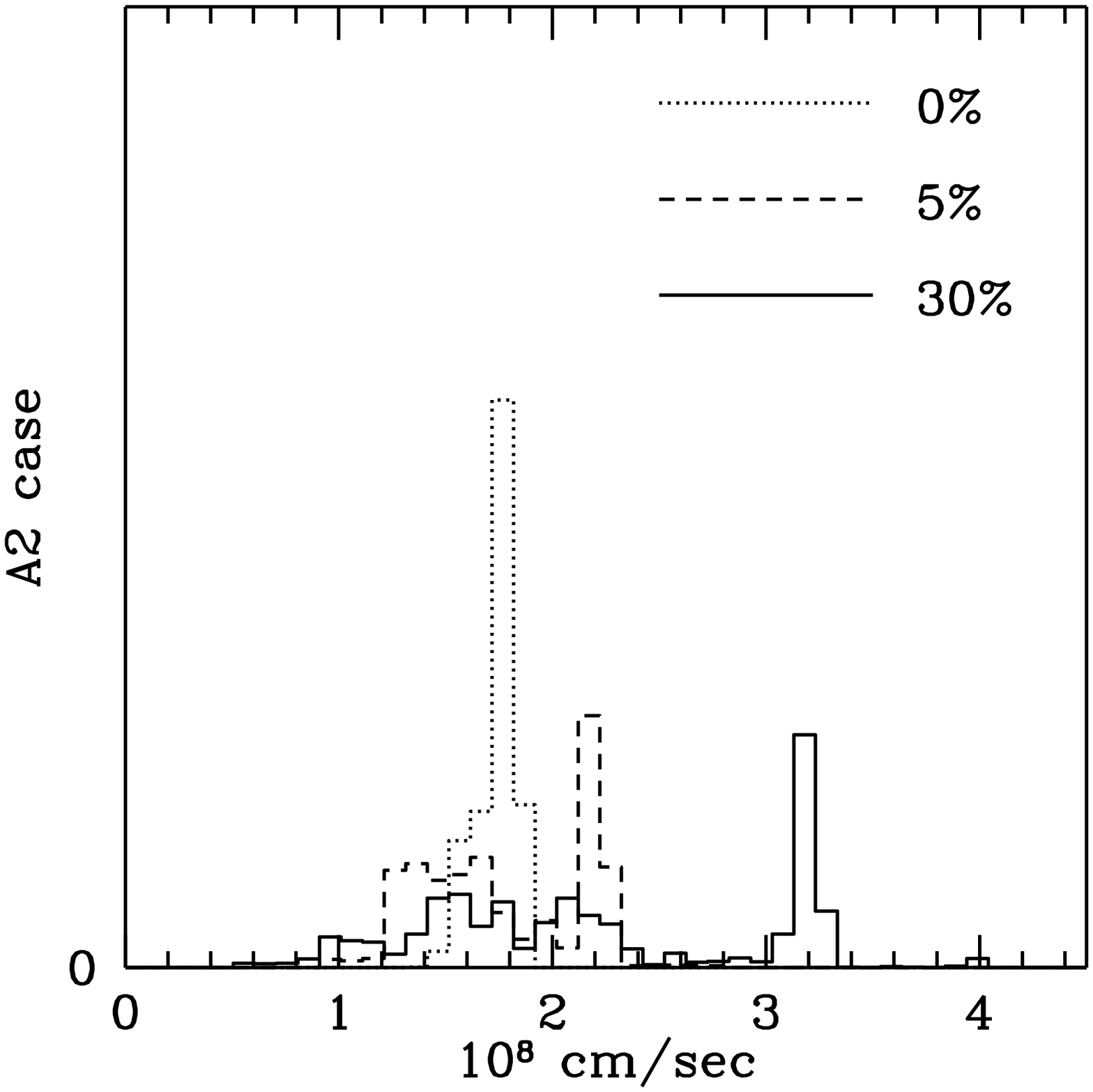}{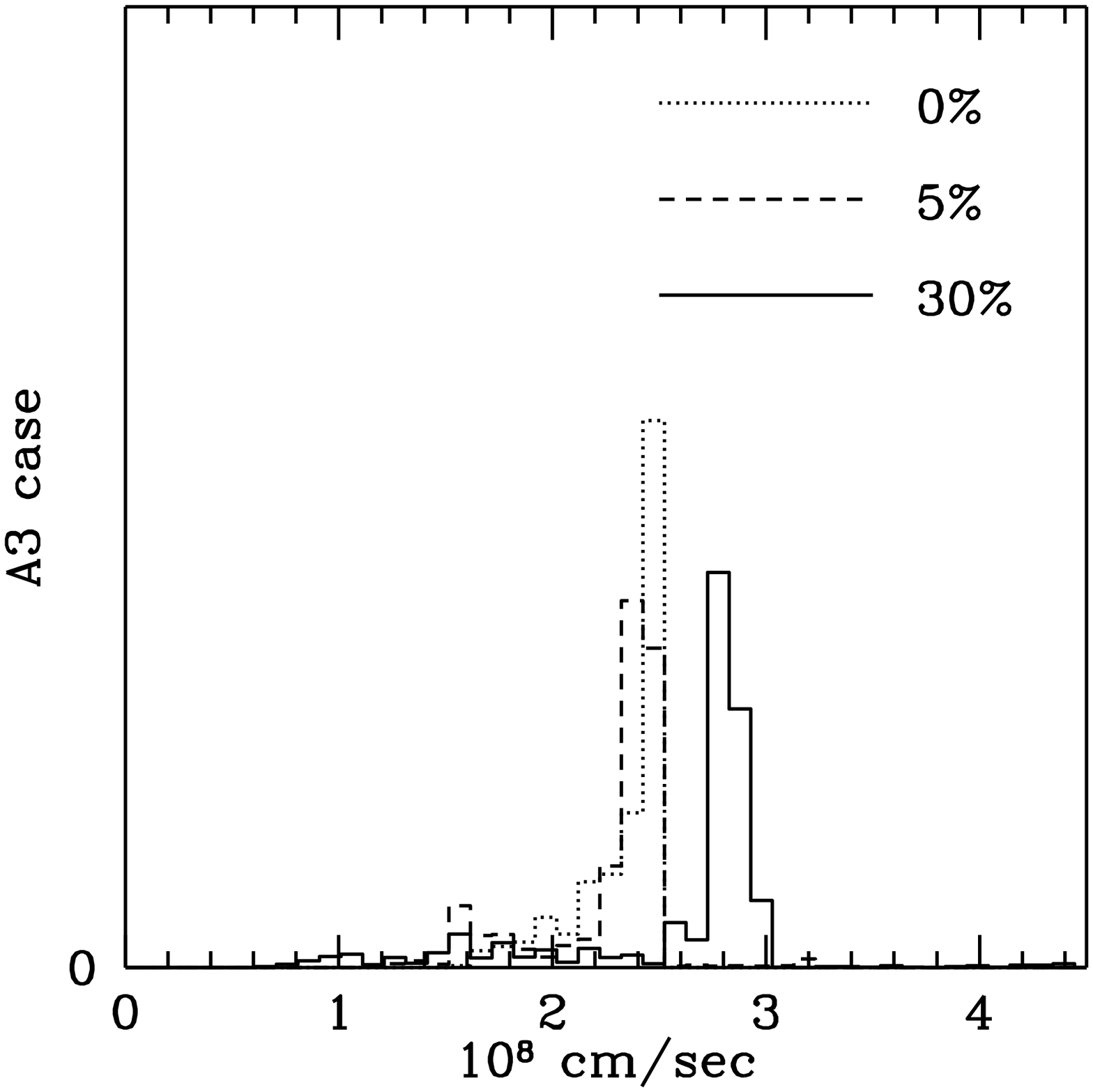}
\figcaption{Same as Fig.6 but for A2 and A3 model.
\label{fig7}}
\end{figure}

At first, we will make investigation into the velocity distributions
of $\rm ^{56}Ni$ for all models.
Figure~\ref{fig6} and Figure~\ref{fig7} are the results of them.
We can see that the range of velocity gets wider as
the initial perturbation gets larger for any explosion model, that is,
irrespective of the value of $\alpha$.
However, high velocity $\rm ^{56} Ni$ of order 3000-4000 km/sec can not 
be produced in the spherical explosion case even if the initial
perturbation is as large as 30$\%$. On the other hand, such high
velocity $\rm ^{56} Ni$ is produced for any axisymmetric explosion for the
case of 30$\%$ initial perturbation. However, high velocity $\rm ^{56}
Ni$ seems to be produced too much for strong axisymmetric explosion
case (see A2c and A3c model in Figure~\ref{fig7}).
This might be a constraint for the degree of axisymmetric explosion.
We will investigate this suggestion more carefully in the following.

Before we go to the further discussion, we make a comment on the
average velocity of $\rm ^{56}Ni$. In the perturbed models, slower velocity
component and higher velocity component appear naturally compared with
no perturbed models since the way of perturbation is monochromatic.
As $\epsilon$ becomes large, its degree also becomes large. In fact, it can be
seen in Figure~\ref{fig4},~\ref{fig5} that the radius of inner most layer 
is smaller in the $30 \%$ models compared with $5 \%$ models. At the
same time, the mixing width is larger in the $30 \%$ models.
In other words, it seems that the ram pressure due to ingoing bubbles
of lighter elements suppressed the bulk velocity of heavy elements
while a part of them are accelerated by the outgoing mushrooms.
The average velocity of $\rm ^{56}Ni$ is determined how much
$\rm ^{56}Ni$ is taken in the mushroom and how much $\rm ^{56}Ni$ is
remained in the inner most region. 
As a consequence, the average velocity of $\rm ^{56}Ni$ does not
necessarily gets higher together with $\epsilon$. 
In fact, we can see in
Figure~\ref{fig6} that the average velocity of $\rm ^{56}Ni$ becomes
lower in the $5 \%$ models compared with no perturbed models.

\begin{figure}
\epsscale{0.8}
\plottwo{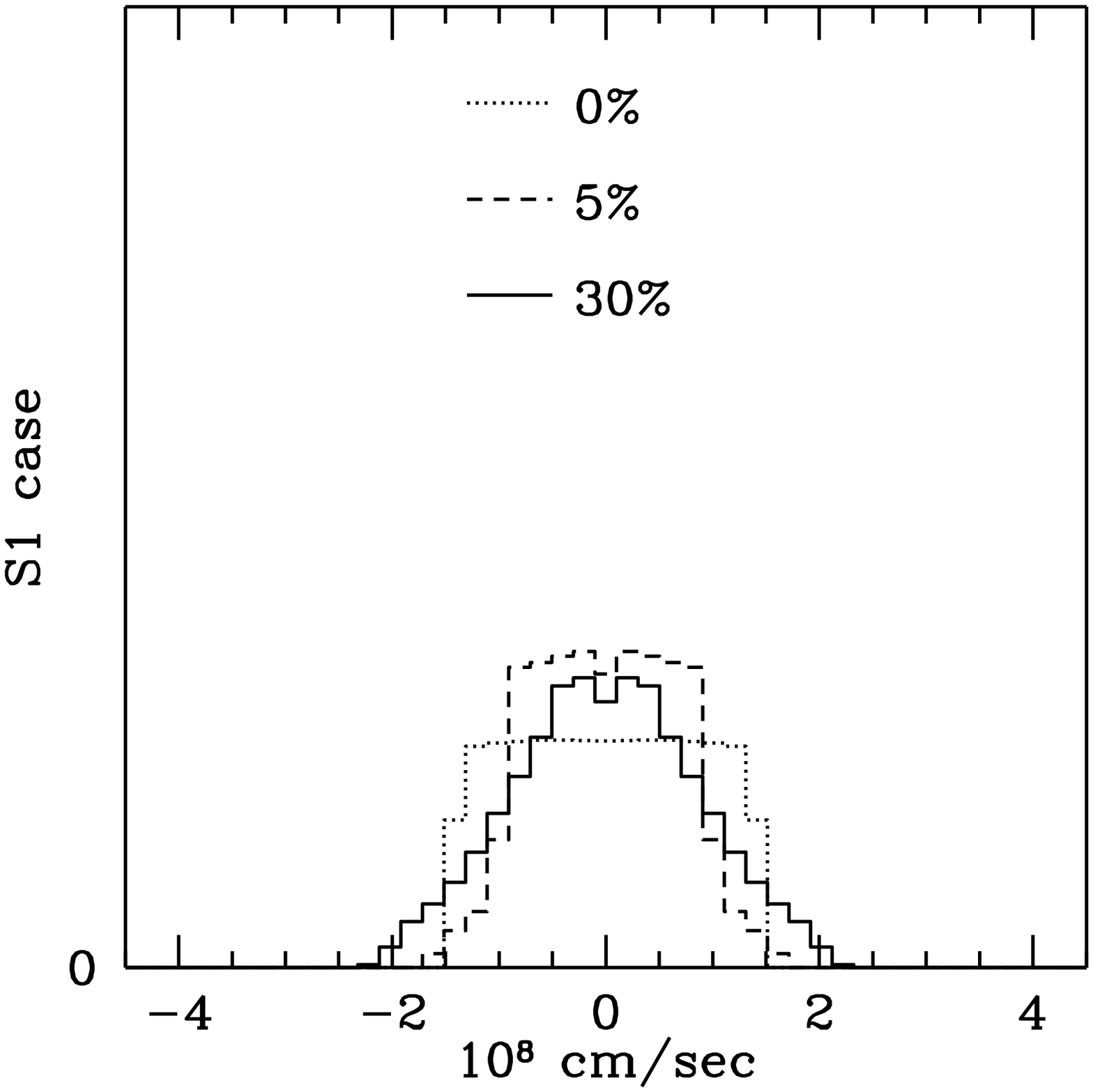}{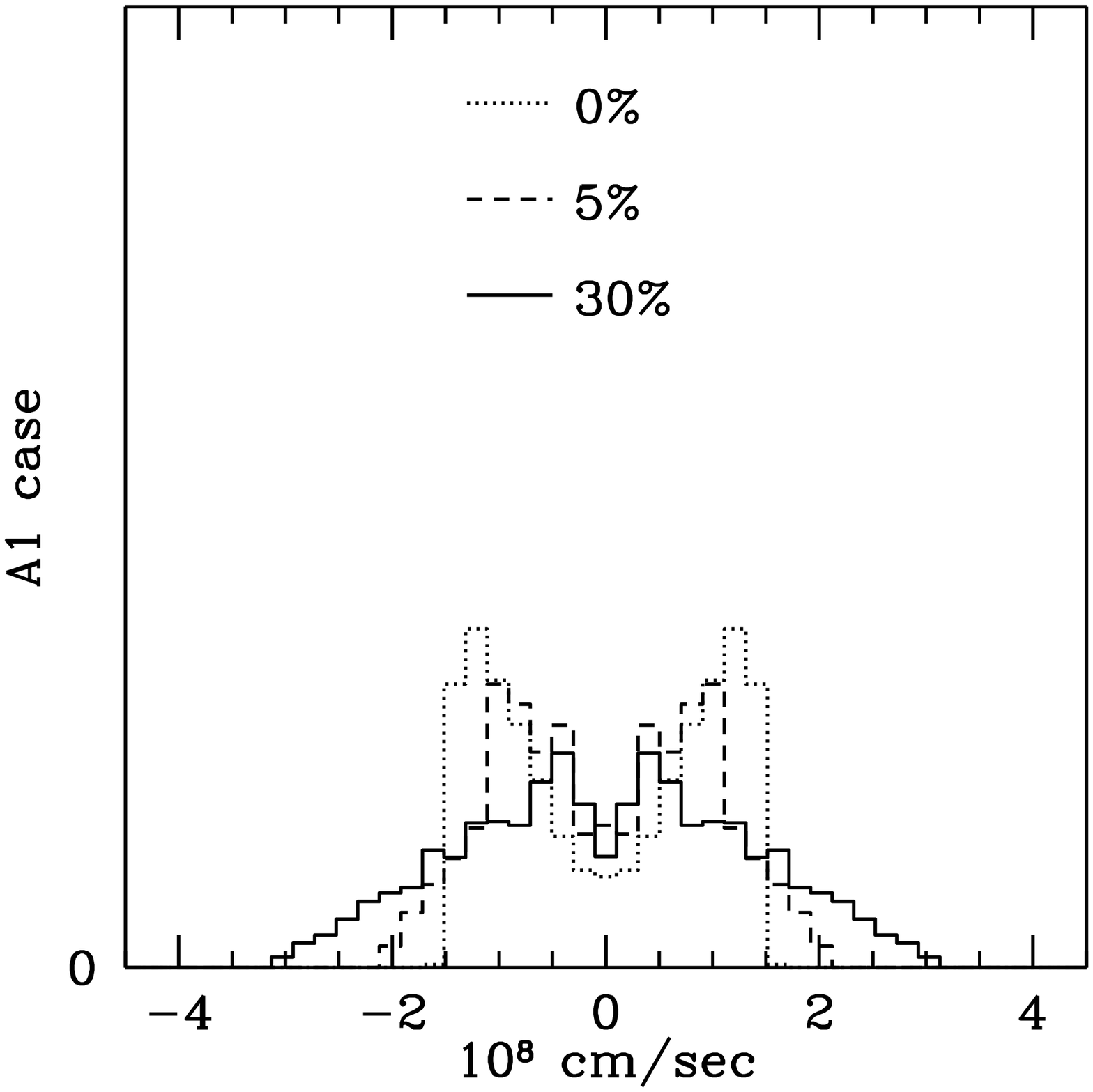}
\figcaption{Line profile of $\rm ^{56}Ni$ seen from $\theta = 44^{\circ}$ at
time = 5000 sec after explosion. Left: S1 model. Right: A1 model. 
\label{fig8}}
\end{figure}

\begin{figure}
\epsscale{0.8}
\plottwo{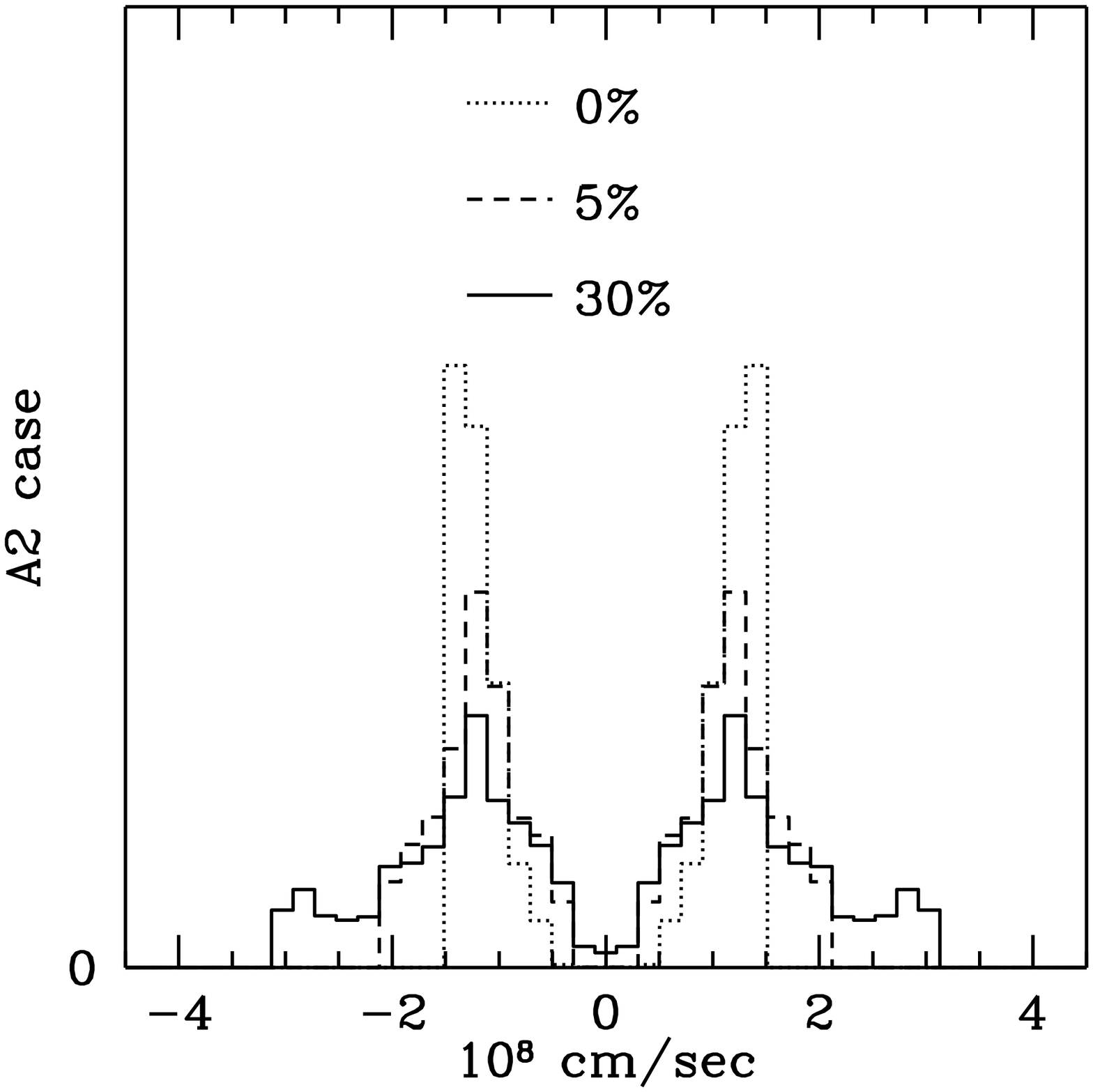}{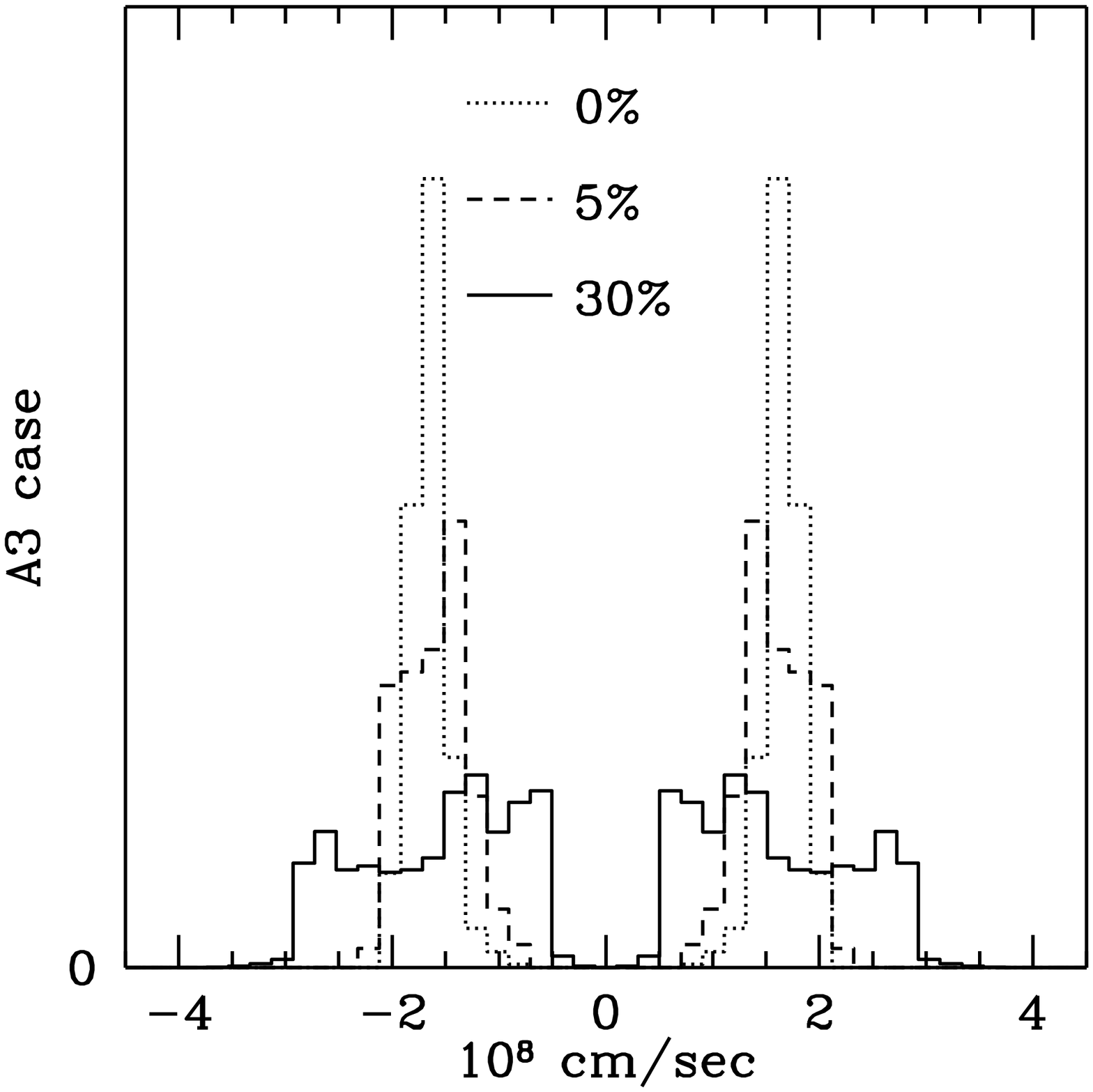}
\figcaption{Same as Fig.8 but for A2 and A3 model.
\label{fig9}}
\end{figure}

Next, we calculate line profiles for optically thin ejecta and compare 
them with observations to examine the suggestion mentioned above.
However, interpretation of calculated line profiles is complicated by
the line of sight of the observer.
The symmetry axis inferred from the observation of the ring of SN
1987A (\cite{plait95}) is different from that inferred
from the speckle interferometry (\cite{papaliolis89}).
In this paper, we 
calculate line profiles seen from $\theta = 44 ^{\circ}$, which is
inferred from the ring of SN 1987A.
The results are shown in Figure~\ref{fig8} and ~\ref{fig9}. For comparison,
observed infrared line profiles of Fe from SN 1987A is shown in
Figure~\ref{fig10}.
As for the spherical explosion, high velocity element can not be seen. 
On the other side, as $\alpha$ gets large, the form of line profile
becomes to be different from that of observations.
In particular, it is unlikely that the strong
axisymmetric explosion model ($\alpha = 3/5, 7/9$ case) can reproduce
the line profile.
We can say that the model A1c most closely resembles
the line profile including high velocity element.

Finally, as mentioned in section \ref{pro},
we must examine the sensitivity of our results on the mass cut.
We can easily guess that the small velocity
element will get larger if the mass cut is
assumed to be spherical since the velocity
is small near the equatorial region, which is cut by the axisymmetric
mass cut. We show line profiles incorporating the spherical
mass cut for comparison in Figure~\ref{fig11}
and ~\ref{fig12}.
As we expected, the figure shows the enhancement of small velocity
element. However, A2c and A3c model are still far from the observed line
profile. From this result, we think A2 and A3 model are rejected by
the observation.
On the other hand, model A1c begins to resemble to the observations.
This supports the proposal that model A1c can reasonably reproduce the 
observation. We also performed a simulation in order to investigate
the effect of asymmetric explosive nucleosynthesis (\cite{nagataki97})
on the velocity distribution of $\rm ^{56}Ni$. It is found, however,
that the effect of asymmetric mass cut is of much more importance.

\placefigure{fig6}
\placefigure{fig7}
\placefigure{fig8}
\placefigure{fig9}
\placefigure{fig10}
\placefigure{fig11}
\placefigure{fig12}

\subsection { Minimum velocity of hydrogen}

\indent

Additionally, we see the minimum velocity of hydrogen. The results are 
summarized in Table \ref{hvel}.
We can identify the tendency for the minimum velocity to become lower
as the degree of axisymmetric explosion gets larger when 
the initial perturbation is set to be zero.
It is also seen that as the initial perturbation amplitude is
increased, the minimum velocity decreases for the same degree of the
axisymmetric explosion.
However, we can not say that the minimum velocity decreases as the degree of
axisymmetric explosion and initial perturbation are both increased.
Although we can not explain the reason clearly, we can only say that
all models which have an initial perturbation larger than $5 \%$, 
can explain the observation, which
shows the minimum velocity of hydrogen $\sim 800$ km/sec.

\placetable{hvel}

\section{ Summary and Discussion} \label{summary}

\indent

We have performed a two-dimensional hydrodynamic calculation 
for an axisymmetric explosion with a periodic perturbation. 
The degree of the explosion axisymmetry is varied and 
three perturbation amplitudes are studied for each explosion. 
We have compared calculated line profiles to observations, with
special attention paid to the high velocity element.
Limits on initial perturbation amplitude and degree of axisymmetry are 
established based on our comparison.

We find the high velocity element can not be produced for any
spherical explosion model even if the initial perturbation amplitude
is as large as $30 \%$.
On the other hand, it is produced by the axisymmetric
explosion, if the initial perturbation amplitude is $30 \%$.
As for the origin of the perturbation, we cannot attribute such a
large perturbation ($\sim 30 \%$) to the structure of the progenitor
(\cite{bazan94}; \cite{bazan97}).
We feel that only the dynamics of the core-collapse
explosion itself can lead such a large perturbation.
However, we must note that if other mechanisms, such as nickel bubbles, works 
effectively, the amplitude to be desired will be less than $30 \%$.
We also add a comment on another possible source of a large
perturbation. It is the offset of the core from the center of mass of
the star. This is being studied now by Peter Goldreich
as the effect of gravity waves
from convective zones acting in constructive interference as they
converge towards the core. This leads to some oscillatory behavior of
the core about the center of mass. However, the amplitude of its
perturbation generated by its effect is not reported yet.

The line profiles also serve as upper limits on the degree of
explosion axisymmetry.
Line profiles are affected by the presence of a mass cut.
However, models of A2c and A3c can not reproduce the observed line 
profile even if different mass cut positions are incorporated.
We think this provides an upper limit of the degree of the
axisymmetric explosion. 
This fact will be important for the simulation of the collapse-driven 
supernova including rotation, magnetic field, and axisymmetric
neutrino radiation.

On the other hand, the weak axisymmetric model, A1c best 
reproduces the observations, including the high velocity element.
Since the line profile is also changed by the angle between our line
of sight and polar axis and radiative transfer, the information
provided by the line profile is ambiguous. We can only say that there is 
an axisymmetric explosion model that can
reproduce the line profile.

To provide an additional constraint, we calculated the lowest velocity
of hydrogen. The
results showed that any explosion model can explain the observed value 
$\sim 800$km/sec if the amplitude of the initial perturbation is
larger $5 \%$.

\begin{figure}
\epsscale{0.8}
\plotone{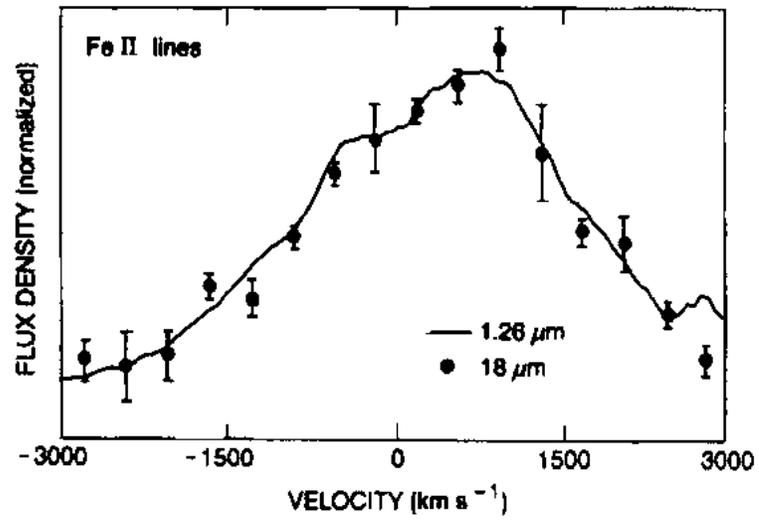}
\figcaption{Observed infrared line profiles for Fe at 1.26 $\mu m$
(solid line, t = 377 days: Spyromilio et al. 1990) and at 18 $\mu m$
(data points, t = 407 days; Haas et al. 1990). Positive velocities
(km/sec) correspond to a redshift.
\label{fig10}}
\end{figure}

\begin{figure}
\epsscale{0.8}
\plotone{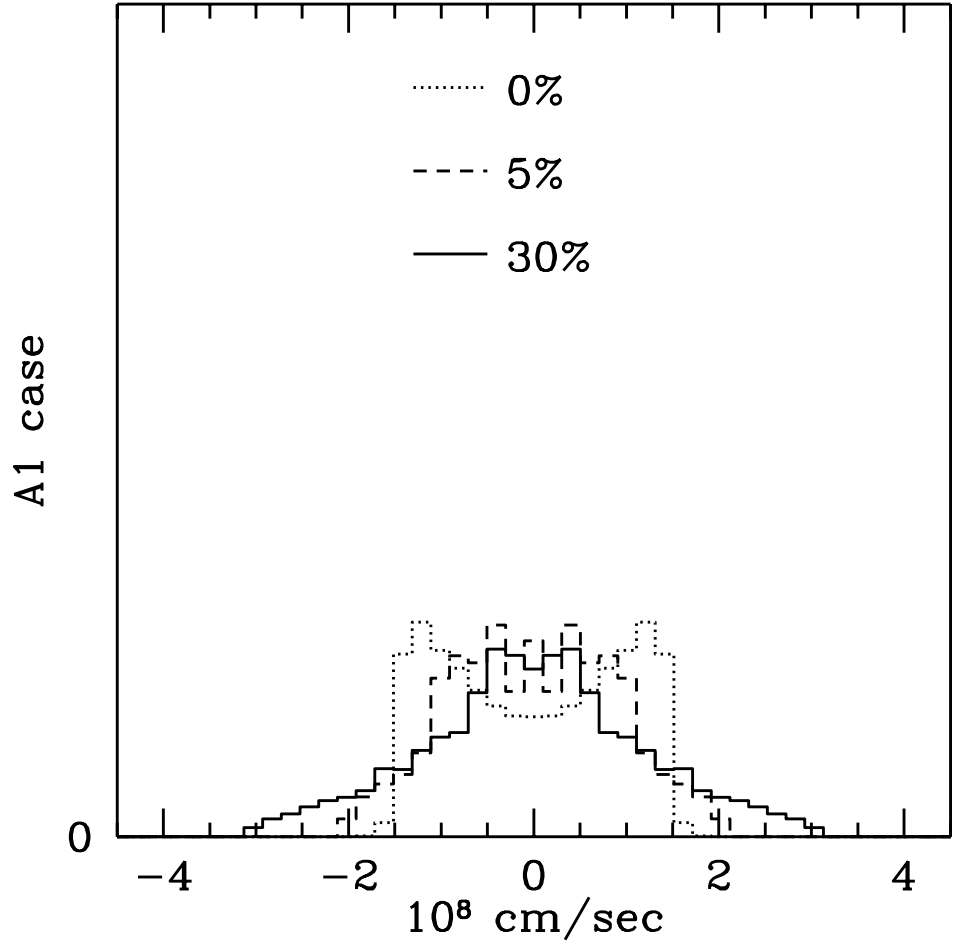}
\figcaption{Line profile of $\rm ^{56}Ni$ for A1 model seen from
$\theta = 44 ^{\circ}$
at time = 5000 sec after explosion under the spherical mass cut. 
\label{fig11}}
\end{figure}

\begin{figure}
\epsscale{0.8}
\plottwo{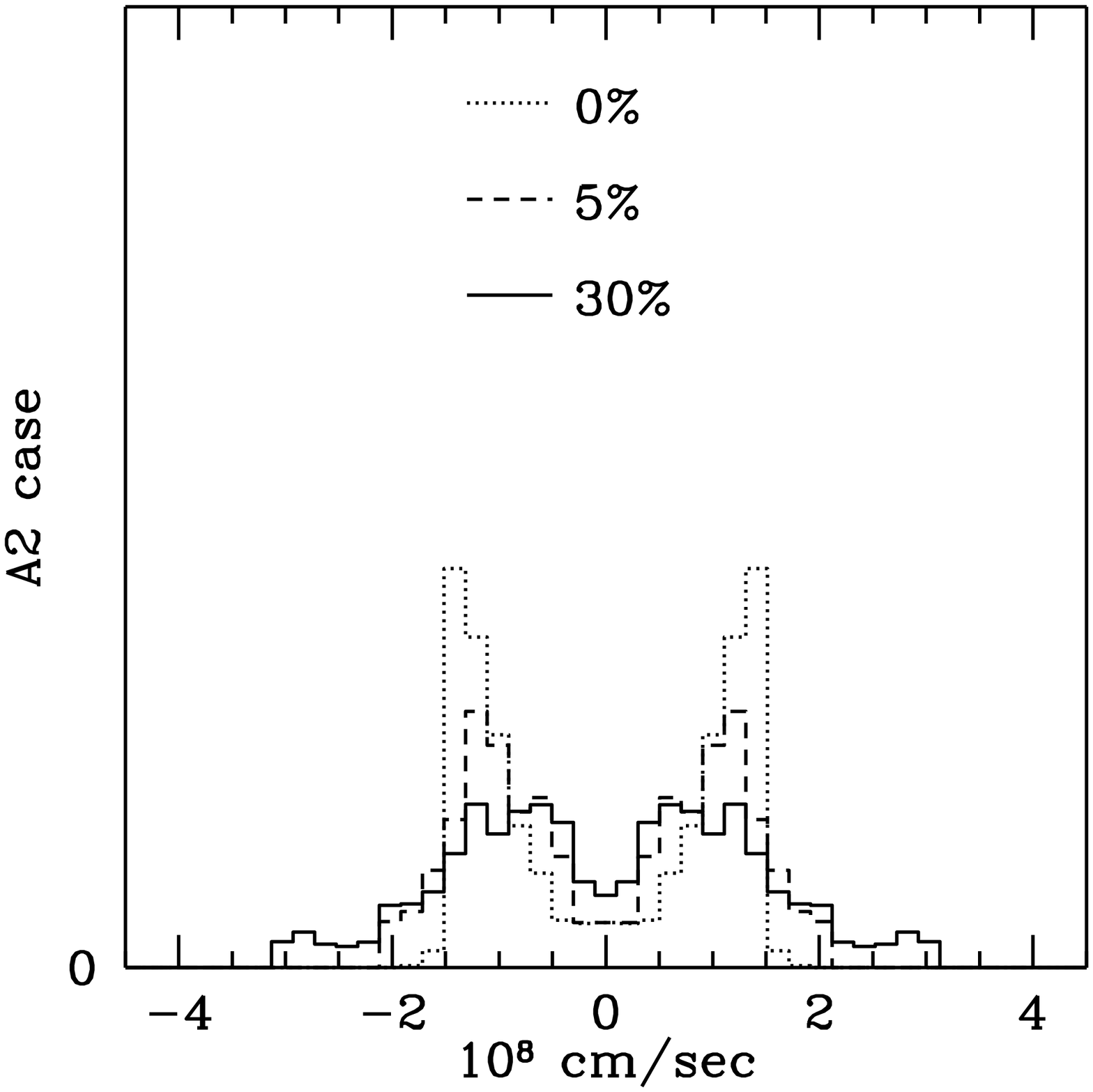}{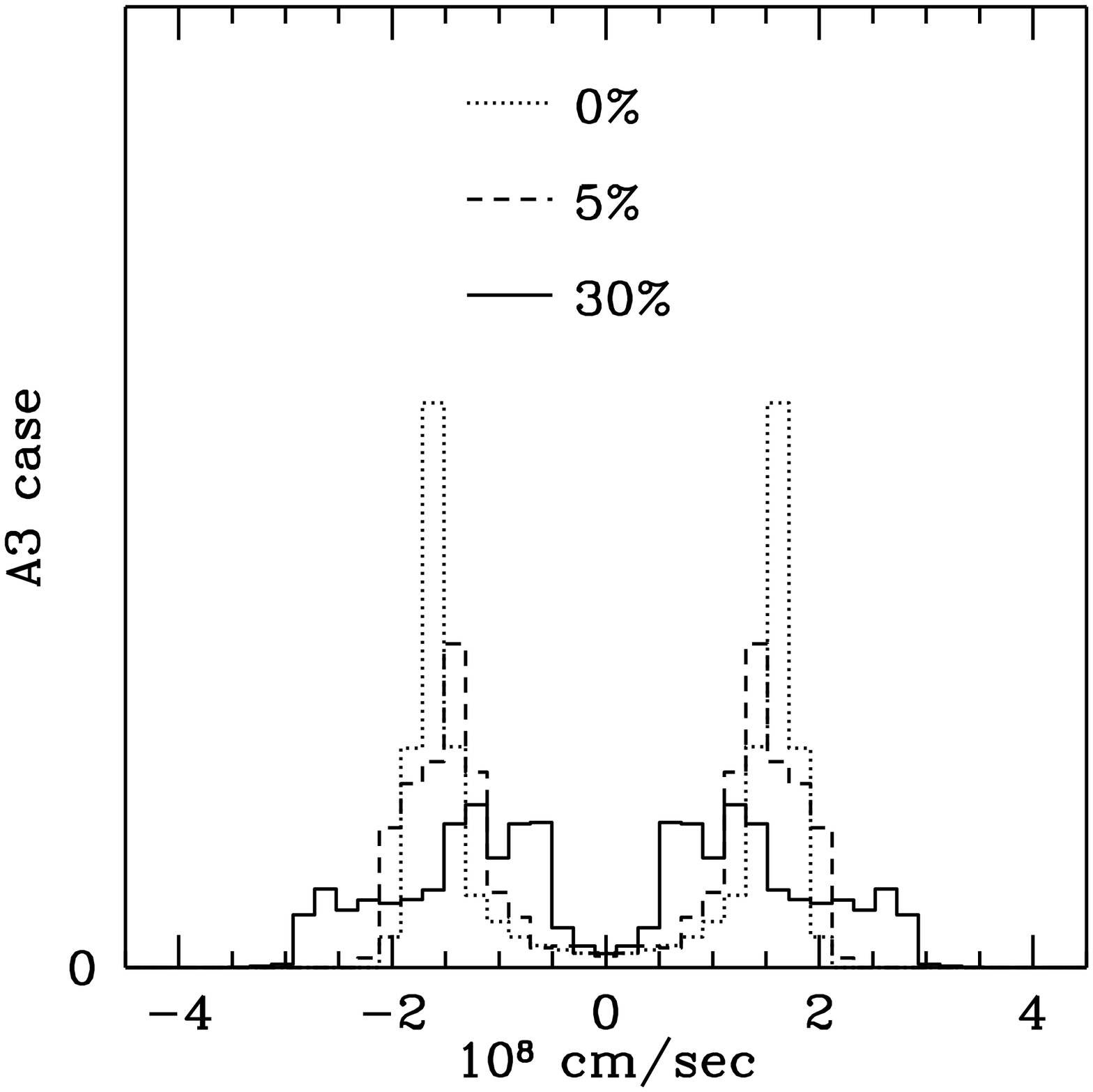}
\figcaption{Same as Fig.11 but for A2 and A3 model.
\label{fig12}}
\end{figure}

We must note that our results are obtained on the assumption of
periodic perturbation method. As stated in
subsection~\ref{perturbetc}, the effect of initial power spectrum of
density perturbations should be explored in future. In particular, it
will be important to perform R-T calculations using the asymmetric
progenitor models (\cite{bazan94}; \cite{bazan97}) since their
two-dimensional models predict the power spectrum of density perturbations. 
Their models may also
have a key for the reproduction of the observed line asymmetry like
that in Figure~\ref{fig10}. 
However, we think that the influence of the power spectrum of the
initial perturbation is relatively small in this study since the
amplitudes of initial perturbations are larger than $5 \%$ of the
expansion speed (\cite{hachisu92}).

We also make a comment on the effect of Coriorlis force. 
When instabilities 
of other targets that have enough angular velocity, such as an
accretion disk around a black hole, are explored (\cite{ruffert94}),   
Coriorlis force will play an important role with respect to the growth 
of instabilities.
However, we neglected the rotation of the mantle and envelope in this paper
since the rotation velocity is much smaller than the explosion
velocity where the R-T instabilities grow.

We will consider the reliability of our calculations.
Grid resolution is relevant since higher resolution can better
reproduce the steep gradient of physical quantities, such as those for density,
pressure, and temperature. Fryxell, M$\rm \ddot{u}$ller, and Arnett
have shown that the minimum resolution required for a 'converged'
model is determined by the hydrodynamic algorithm and that there is a
possibility that numerical errors might dominate the physical
instability in much work which has already been done.
On the other hand, Hachisu et al have shown the mixing
width is
insensitive to resolution in their Roe's third order TVD 
scheme if the initial amplitude of the velocity perturbation is larger
than $1 \%$ of the local sound speed (\cite{hachisu92}). 
It is important to see the structure of the mushrooms in detail, to be 
sure, but our main purpose is to see the global behavior of the
material. We think our calculations have enough resolution for that 
purpose. 
For example, we can see the
mushroom-like 'Finger' in the S1b case. Since the finger cannot be seen
clearly in \cite{yamada91} under the same condition but can be seen in other
groups (e.g., \cite{hachisu92}), we conclude that our method is an
improvement to that in
\cite{yamada91} and that our axisymmetric explosion models employ
sufficient resolution to accurately describe the global behavior of
the material.

\acknowledgements
S. N. is grateful to S. Yamada for a
useful discussion. We thank D. Arnett for his kind comments and review 
on this manuscript. S. N. and T. M. S. are a JRA (Junior Reserch
Associate) member and a Special Postdoctral Researcher in The
Institute of Physical and Chemical Research (RIKEN), respectively. 
This work is supported part by the Grant-in-Aid by the
Ministry of Education, Science and Culture of Japan.(No. 07CE2002,
07640386 and 07304033).

\vskip1.0cm


\clearpage

\begin{deluxetable}{lc}
\footnotesize
\tablecaption{Initial positions of test particles \label{position}}
\tablewidth{0pt}
\tablehead{
\colhead{radius [cm]} & \colhead{number}} 
\startdata
$ (1.5-3.0) \times 10^8$  & 1000  \nl
$(3.0-9.0) \times 10^8$    & 1000  \nl
$9.0\times 10^8 - 6.3 \times 10^9$   & 1000  \nl
$6.3\times 10^9 - 4.8 \times 10^{10}$  & 1000  \nl
$4.8\times 10^{10} - 3.3 \times 10^{12}$      & 1000  \nl
\enddata 
\end{deluxetable}

\begin{table*}
\begin{center}
\begin{tabular}{c|rrrrrrrrrrr}
\tableline
\tableline
Model       & S1  &  A1  &   A2 &  A3  \\
$\alpha$    &  0  & 1/3  &  3/5 &  7/9 \\
$V_p$:$V_e$ & 1:1 & 2:1  &  4:1 &  8:1 \\
\tableline
\end{tabular}
\end{center}

\tablenum{2}
\caption{
Models for the initial shock wave. The first row
shows names for each model. Second is the value of $\alpha$ for each model.
Third is the ratio of the velocity in
the polar region ($\theta = 0 ^{\circ}$) to that in the equatorial
region ($\theta = 90 ^{\circ}$).  \label{model}}

\end{table*}

\begin{table*}
\begin{center}
\begin{tabular}{ccccccccccccccc}
\tableline
\tableline
 Model    & $\alpha$   &  Perturbation[$\%$]   &   Model   &  $\alpha$
& Perturbation[$\%$]  \\
\tableline
S1a & 0   & $ 0 $ & A2a & 3/5 & 0      \\
S1b & 0   & $ 5 $ & A2b & 3/5 & 5      \\
S1c & 0   & $ 30$ & A2c & 3/5 & 30     \\
A1a & 1/3 & $ 0 $ & A3a & 7/9 & 0      \\
A1b & 1/3 & $ 5 $ & A3b & 7/9 & 5      \\
A1c & 1/3 & $ 30$ & A3c & 7/9 & 30     \\
\tableline
\end{tabular}
\end{center}

\tablenum{3}
\caption{
Name of each model, the value of $\alpha$, and the amplitude of the
initial perturbation.
\label{models}}

\end{table*}

\begin{table*}
\begin{center}
\begin{tabular}{c|rrrrrrrrrrr}
\tableline
\tableline
     & S1    &  A1   &   A2   &  A3   \\
\tableline
 a   &  1.6  &  1.5  &   1.3  &  1.1  \\
 b   &  0.87 &  0.89 &   0.86 &  0.90 \\
 c   &  0.85 &  0.53 &   0.57 &  0.76 \\
\tableline
\end{tabular}
\end{center}

\tablenum{4}
\caption{
Minimum velocity of hydrogen at time = 5000 sec. The unit is
$10^{8}$cm/sec. 'a' denotes that the amplitude of initial perturbation
= 0$\%$. 'b' and 'c' denote $5 \%$ and $30 \%$, respectively. 
\label{hvel}}

\end{table*}

\end{document}